\documentclass{aa}
\usepackage{natbib}
\usepackage[varg]{txfonts}
\usepackage{graphicx}
\usepackage{float}
\usepackage{bm,bbm,amssymb,color, amsmath, color}
\usepackage{enumerate}
\usepackage{url}
\usepackage[section]{placeins}

\usepackage{hyperref}

\usepackage{etoolbox}
\makeatletter
\patchcmd{\@makecaption}
  {\scshape}
  {}
  {}
  {}
\makeatother

\begin{document}

\title{Rapid elimination of small dust grains in molecular clouds }

\author{Kedron Silsbee\inst{1} \and Alexei V. Ivlev\inst{1} \and Olli Sipil\"a\inst{1} \and Paola Caselli\inst{1}  \and Bo Zhao\inst{1} }
\institute{Max-Planck-Institut f\"ur Extraterrestrische Physik, 85748 Garching, Germany \email{ksilsbee@mpe.mpg.de}} 
\abstract {We argue that impact velocities between dust grains with sizes of less than $\sim 0.1$ $\mu m$ in molecular cloud cores are dominated by drift arising from ambipolar diffusion.  This effect is due to the size dependence of the dust coupling to the magnetic field and the neutral gas.  Assuming perfect sticking in collisions up to $\approx 50$ m/s, we show that this effect causes rapid depletion of small grains,  consistent with starlight extinction and IR and microwave emission measurements, both in the core center ($n \sim 10^{6}$ cm$^{-3}$) and envelope ($n \sim 10^{4}$ cm$^{-3}$).  The upper end of the size distribution does not change significantly if only velocities arising from this effect are considered.  We consider the impact of an evolved grain-size distribution on the gas temperature, and argue that if the depletion of small dust grains occurs as  expected from our model, then the cosmic ray ionization rate must be well below $10^{-16}$ s$^{-1}$ at a number density of $10^{5}$ cm$^{-3}$.}


\titlerunning{Rapid Elimination of Small Dust Grains in Molecular Clouds}
\authorrunning{K. Silsbee et al.}
\maketitle

\section{Introduction}
Knowledge of the dust grain-size distribution is crucial to interpretating observations in molecular clouds.  There is evidence that the size distribution in dense regions contains fewer small grains \citep{Weingartner01} than the dust in the diffuse ISM and some grains of at least micron size \citep{Pagani10}.  There has been theoretical work modeling  the coagulation process.  \citet{Ossenkopf93} made estimates for the contribution of several processes (e.g., turbulence, Brownian motion, forces arising from grain asymmetries and gravitational settling), to the relative velocity between dust grains, and determined turbulent motions to be dominant at densities $\leq 10^8$ cm$^{-3}$.  He ran coagulation simulations for regions with densities between $10^5$ and $10^9$ cm$^{-3}$.  It was found that the optical properties of the dust distribution change significantly due to the removal of small particles, but the upper edge of the size distribution does not change appreciably within $10^5$ years for a density of $10^6$ cm$^{-3}$.
\par
\citet{Ormel09} used a detailed model for the outcome of grain-grain collisions, in which both grain size and structure are altered.  They assumed collision velocities to be due to gas turbulence, as this was argued to be dominant in \citet{Ossenkopf93}, and they ignored the effect of grain charge on the collisional rates.  A great deal of uncertainty exists regarding the lifetimes of molecular cloud cores \citep{Lee99, Ciolek00}, although a few million years seems to be a good estimate based on statistical studies by  \citet{Konyves15}, among others.  The lifetime of the central regions of a dense core (with densities above $10^5$ cm$^{-3}$, as in prestellar cores; e.g., \citealt{Crapsi05}),  similar to that modeled in this paper, is likely somewhat shorter.  \citet{Ormel09} ran their standard simulation for $5 \times 10^7$ years with a gas density of $10^5$ cm$^{-3}$.  In this longer time they found that large aggregates (tens of $\mu$m)  form, and fragmentation becomes important.
\par
Recent non-ideal magnetohydrodynamic simulations have pointed out the importance of the removal of very small grains for the formation of protoplanetary disks \citep{Zhao16, Zhao18a}. It is therefore important to look into this problem quantitatively and for physical conditions relevant for stellar system formation. In particular, we need to determine the  timescale for very small grain removal in prestellar cores, which represent the initial conditions in the process of star and planet formation. If this depletion timescale is significantly shorter than the dynamical timescale, then the diffusion of magnetic fields (by ambipolar diffusion and the Hall effect) becomes efficient during the protostellar collapse, which greatly promotes the formation of protoplanetary disks.  The depletion of very small grains also affects the chemical evolution of prestellar and protostellar cores. Very small grains  provide a large area on which surface chemical processes can proceed and can also  become the main carriers of electrons, thus affecting ion-molecule chemistry in the gas phase.
\par
In this work, in addition to turbulence, we study the effect of a different source of relative velocity between grains.  \citet{Ciolek96} considered the role of ambipolar diffusion in changing the spatial distribution of grains.  In a collapsing cloud, the large grains are pulled to the center of the cloud by the infalling neutral material, whereas the small grains (mostly negatively charged)  stay coupled to the magnetic field.  In this paper we note that this differential motion leads to a substantial additional source of collision velocity between small and large grains, which operates even in the absence of turbulence.  This is particularly noteworthy given evidence \citep[e.g.,][]{Fuller92, Pineda10} that turbulent velocities in dense regions are substantially subsonic.

\section{Cloud model}
We take the cloud density profile that \citet{Tafalla02} used to model the prestellar core L1544.  This is a spherically symmetric cloud with the ${\rm H_2}$ number density given by 
\begin{equation}
n(r) = \frac{n_c}{1+(r/r_0)^{2.5}}.
\label{TafallaDensity}
\end{equation}
Here $n_c = 1.4 \times 10^6$ cm$^{-3}$ is the central density of the core, and $r_0 = 0.014$ pc = 2,900 AU is the rough scale of the density peak.  In this paper we focus on two representative environments within the cloud.  First we consider the location at $r = r_0$, taken to be representative of the environment near the cloud center, but far enough from the center that motions due to ambipolar diffusion and gravitational settling are still present.  We denote this our ``core center'' location.  We also consider a location at $r = 7.2 r_0$ where $n = 10^4$ cm$^{-3}$, denoted our ``envelope'' location. In what follows we assume that the primary ion is ${\rm H_3^+}$ in the core center \citep{Caselli03} and ${\rm HCO^+}$ \citep{Caselli02} in the envelope.  We assume a composition with a helium-to-hydrogen atom number ratio of 10\% and ignore all other species.  From the temperature profile shown in the top panel of Figure 3 in \citet{Keto10} we take the temperature to be 6.5K in the core center and 10K in the envelope, consistent with gas measurements done by \citet{Crapsi07}. We take the ion density from \citet{Caselli02} (their model 3), but scale it with the square root of the cosmic ray (CR) rate $\zeta$:
\begin{equation}
n_i = 2.1 \times 10^{-5} \left(\frac{n_{\rm H_2}}{\rm 1\, cm^{-3}}\right)^{-0.56} \left(\frac{\zeta}{10^{-16}\, {\rm s^{-1}}}\right)^{0.5}.
\label{eq:ionFrac}
\end{equation}
The value of $\zeta$ is rather uncertain, but thought to be a decreasing function of the column density $N$.  We take the model for $\zeta$ given in \citet{Padovani18} (assuming their high proton spectrum), which yields $\zeta = 10^{-16}$ s$^{-1}$ in the core center and $\zeta = 3 \times 10^{-16}$ s$^{-1}$ in the envelope.  This is somewhat higher than previous estimates in the literature.  Their model however gives the correct ionization rate at lower densities where $\zeta$ can be determined from ${\rm H_3^{+}}$ observations \citep{Neufeld17}.  A summary of the important parameters in each environment is provided in Table 1, with references provided at the bottom.
\begin{table}[h!]
  \begin{center}
    \caption{Parameters of the two characteristic environments}
    \label{tab:table1}
    \small
    \begin{tabular}{l c c c c } 
      \textbf{Variable} & \textbf{Symbol}& \textbf{Units} & \textbf{Core value} & \textbf{Envelope value} \\
      Radial distance & $r$ & AU & $2.9 \times 10^3$ & $2.1 \times 10^4$\\
      H$_2$ density$^1$ & $n$ & cm$^{-3}$ & $7 \times 10^5$ & $10^4$\\ 
      Temperature$^2$ & $T$ & K & 6.5 & 10.0 \\
      Sound speed & $c_s$ & m/s & 152 & 188\\
      CR rate$^3$ & $\zeta$ & s$^{-1}$ & $1.0 \times 10^{-16}$ & $3.0 \times 10^{-16}$\\
      Ion fraction$^4$ & $n_i/n_{\rm H_2}$ & & $1.1 \times 10^{-8}$ & $2.1 \times 10^{-7}$\\
     
    \end{tabular}
  \end{center}
  \begin{flushleft}
{\tiny
1. \citet{Tafalla02} \\
2. \citet{Keto10}\\
3. \citet{Padovani18} \\
4. \citet{Caselli02} \\
}
\end{flushleft}
\vspace{-.5cm}
\end{table}

\section{Collision rates and velocities}
We consider both random and systematic grain motions.  Random motions are due to Brownian motion (important for the smallest grains) and motions due to the coupling with gas turbulence, important for larger grains.  Systematic grain motions arise because grains of different sizes and charge states couple differently to the neutral gas and to the magnetic and gravitational fields.
\subsection{Grain charging}
\label{sect:grainCharge}
As we  show below,  the collision cross sections and the grain dynamics are both affected by the grain charge.  We used the grain charging model from \citet{Ivlev15}.  This includes a combination of plasma charging (including the effect of the charge-induced dipole interaction on the plasma charging rates), photoelectric charging from the UV photons produced by the interaction of CRs and ${\rm H_2}$ molecules \citep{Prasad83}, and additional negative charging from collisions of dust with CR electrons in the low-energy tail of the spectrum.  However, instead of calculating the local electron spectra ourselves, we estimated them at the three column densities shown in Figure 2 from \citet{Ivlev15} and interpolated between them.  This will not affect the accuracy of our results since, as demonstrated in \citet{Ivlev15}, direct charging from accumulating low energy CRs has  a very small effect relative to the other charging mechanisms.
\par
Table 2 shows the charge states of the grains in the envelope.  Each row corresponds to a different grain size.  The triples of numbers comprising each entry of the table are the fractions of grains that are negatively charged, neutral, and positively charged.  The three columns are for different ionization rates $\zeta$ in units of $\zeta_{\rm P18}$, the value shown in Table 1 for the envelope.
 \begin{table}[h!]
        \caption{Fraction of negative, neutral, and positive grains at three different ionization rates.}
    \label{tab:table1}

    \begin{tabular}{c c c c} 
      a (nm) & $\zeta = 0.1 \zeta_{\rm P18}$ & $\zeta = \zeta_{\rm P18}$ & $\zeta = 10 \zeta_{\rm P18}$ \\
      \hline
      1 & $[\overset{-}{0.77},\overset{0}{0.23}, \overset{+}{0.00}]$ & $[\overset{-}{0.75}, \overset{0}{0.25}, \overset{+}{0.00}]$ & $[\overset{-}{0.70}, \overset{0}{0.30}, \overset{+}{0.00}]$\\
      10 & [0.88, 0.12, 0.00] & [0.81, 0.19, 0.00] & [0.63, 0.36, 0.01]\\
      100 & [0.86, 0.14, 0.00] & [0.67, 0.30, 0.03]&[0.36, 0.50, 0.14]\\
      1000 & [0.70, 0.23, 0.07] & [0.30, 0.30, 0.40]&[0.02, 0.05, 0.93]\\
    \end{tabular}

\end{table} 
The charge distribution is dominated by grains of a single charge except at high CR rates, where large grains can develop multiple positive charges.  The charge distribution in the core center is dominated by plasma charging, and consists almost exclusively of grains with either 0 or -1 charge.
 \subsection{Brownian motion}
We assume that each dust grain has a thermal velocity distribution, which leads to expected collision velocities
\begin{equation}
\sqrt{\langle \Delta V_{\rm 12, BM}^2\rangle} = \sqrt{\frac{8 k_BT}{\pi \mu}},
\end{equation}
where $\mu = m_1m_2/(m_1 + m_2)$ is the reduced mass of the two dust grains.

\subsection{Motion due to coupling with gas turbulence}
\label{turbulence}
If transonic turbulence is present, motions of dust grains arising in response to this turbulence likely dominate the collision velocities between the larger grains \citep{Ossenkopf93}.  To calculate these motions, we use the results from \citet{Ormel07}.  We assume a Kolmogorov turbulent energy spectrum with no smallest scale (i.e., $E(k) \propto k^{-5/3}$ for all $k$ greater than $k_L$).  Here $k$ is the spatial frequency, and $k_L$ is the $k$ corresponding to the turbulent injection scale.  
\par
We take the rms collision velocity from \citet{Ormel07}.  We use their equations (5)-(10) in the limit of infinite Reynolds number.  These equations specify the collision velocity as a function of the turbulent velocity $v_g$, and the Stokes numbers of the two grains, defined as $St = \tau_s/\tau_L$, where $\tau_L$ is the turbulent crossing time at the injection scale and $\tau_s$ is the stopping time of the grain, given by

\begin{equation}
\tau_s = \frac{3 m}{4 v_{\rm th}^* \rho_g \sigma_g},
\label{eq:ts}
\end{equation}
where $v_{\rm th}^* = 0.92 \sqrt{4k_BT/(\pi m_p)}$ is the thermal velocity scale of the gas (with a factor to account for the helium fraction), $\rho_g = 2.8 m_p n$ is the gas density, and $\sigma_g = \pi a^2$ (where $a$ is the grain radius) is the collision cross section between grains and gas particles.  
\par
This treatment likely overestimates the collision velocities between smaller dust grains for two reasons.  First there must be a  cutoff at some $k$ due to either ion-neutral friction or neutral viscosity \citep{Xu16}.  In addition, even if the cutoff is not sharp, there is evidence from simulations \citep{Downes12} that the spectrum is steeper than Kolmogorov in realistic ISM conditions.  For this reason, we consider models both with and without turbulence, and focus much of our paper on the previously ignored but actually better constrained source of relative velocity discussed in the following section.

\subsection{Systematic drifts}
\label{sect:systematicDrifts}
In this section we consider systematic drifts due to the size dependence of the grain coupling to the neutral gas, and gravitational and magnetic fields.
\par
\citet{Ossenkopf93} considered the differential drift of grains in a gravitational field and found this effect to be subdominant to the turbulent motions.  He was, however, considering turbulence that was sonic at the Jeans length, and there is evidence \citep[e.g.,][]{Fuller92, Lada08, Pineda10} that turbulent velocities in dense cores are much lower.  In this regime, gravitational settling can be the dominant source of collision velocity for grains larger than a few tenths of a micron.  However, as we show later, this is unlikely to have a significant impact on the size distribution. 
\par
 \citet{Ciolek96} considered the drift of small dust grains relative to large ones as a result of their coupling to the magnetic field.  To our knowledge this motion has not been considered in studies of coagulation.  In this section we show that this effect can be the dominant source of relative motion for grains up to $\sim 100$ nm, even when significant turbulence is considered.
\par
To estimate the drift rate of the dust grains as a result of coupling to the magnetic and gravitational fields as well as gas drag, we assume that the magnetic field is able to keep the ions stationary in the frame of the cloud center.  We then assume that the motion of the neutrals can be determined by balancing the drag force from collisions with ions against the gravitational force.  Considering both the molecular hydrogen and the helium, the drift velocity $v_{\rm in}$ between the ions and neutrals is determined by
\begin{equation}
c_d g \rho_g = n_i v_{\rm in} \left[\mu_{\rm H_2} n_{\rm H_2}  \langle \sigma v \rangle_{\rm H_2}  + \mu_{\rm He} n_{\rm He}   \langle \sigma v \rangle_{\rm He} \right],
\label{eq:forceBalance}
\end{equation}
 where $g$ is the local gravitational field;  $\mu_{\rm H_2}$ and $\mu_{\rm He}$ are the reduced masses of ${\rm H_2}$ and helium, respectively, in collisions with the ions; and  $\langle \sigma v \rangle$ is the ion-neutral momentum transfer cross section averaged over a Maxwellian velocity distribution.  Following \citet{Raizer11}, $\langle \sigma v \rangle$ is given by $4 \pi a_0^2 \sqrt{(\alpha/a_{0}^{3})I_H/\mu}$, where $a_0$ is the Bohr radius, $I_H$ the Rydberg energy, and $\alpha$ is the polarizability of the neutral species. The value of   $\alpha/a_0^3$ is 5.52 for ${\rm H_2}$ and 1.39 for He.  We   added a fudge factor $c_d \in [0, 1]$ related to the fact that the field is not purely azimuthal; the ions may be to some extent dragged in along with the neutrals, or the cloud may be supported by gas pressure in addition to the magnetic field.  Solving Equation \eqref{eq:forceBalance} (with $c_d = 1$) for $v_{\rm in}$ yields $v_{\rm in} = 5.2 \times 10^3$ cm/s in the core center and $4.2 \times 10^3$ cm/s in the envelope.
 \begin{figure}
\centering
\includegraphics[width=1.0\columnwidth]{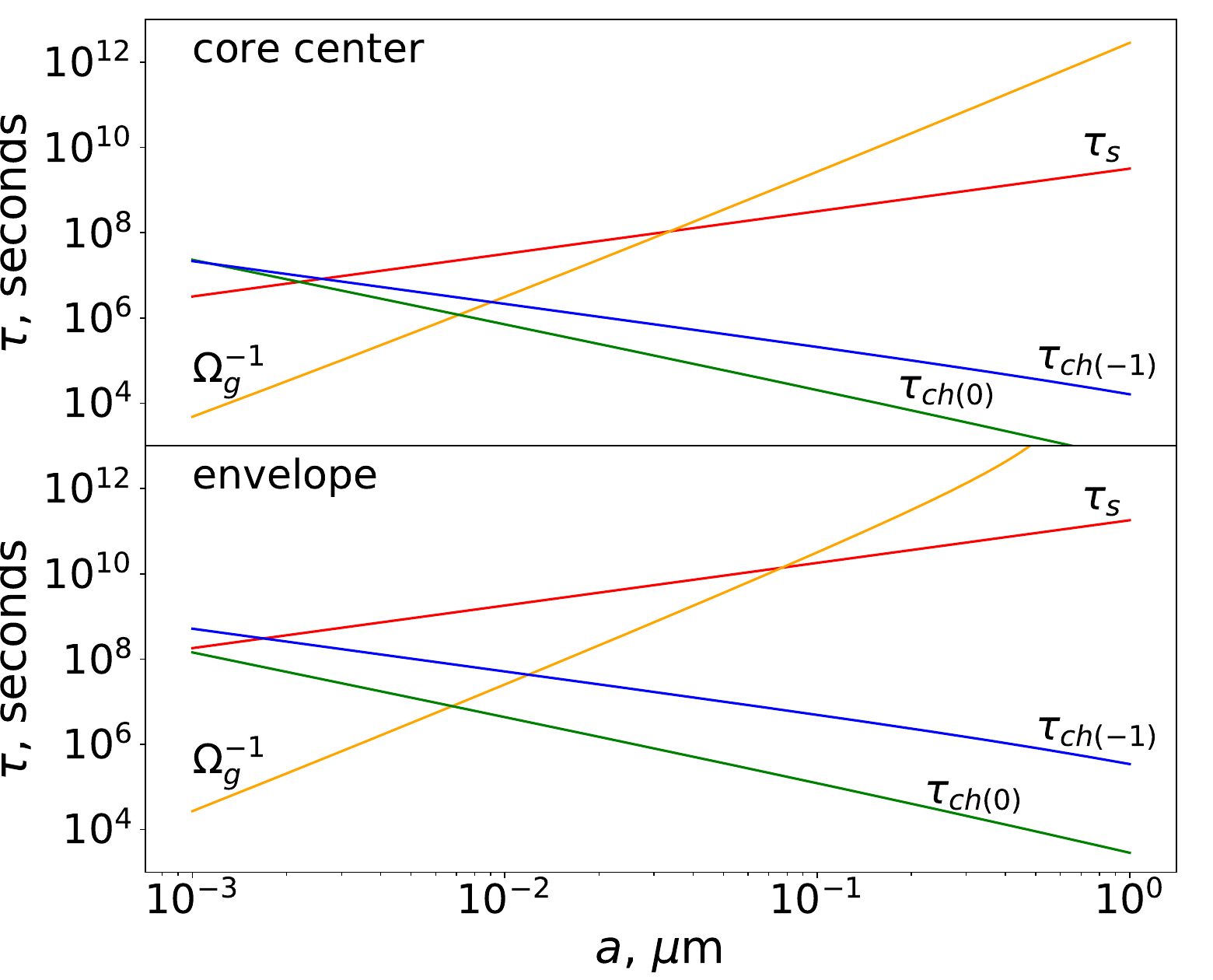}
\caption{Timescales for processes relevant to grain dynamics:  $\tau_s$ is the stopping time of the grain, given by Equation \eqref{eq:ts};  $\Omega_g^{-1}$ is the inverse grain gyrofrequency given by Equation \eqref{eq:Omegag};  $\tau_{ch(0)}$ is the time for a neutral grain to acquire a negative charge; and $\tau_{ch(-1)}$ is the time for a grain of charge -1 to become neutral.  The top panel is for the core center, and the bottom is for the envelope.}
\label{fig:timescales}
\end{figure}

In order to determine the grain dynamics, we must understand how the charging timescales compare with the other two relevant dynamical timescales in the problem:  the stopping time $\tau_s$ and the inverse of the grain gyrofrequency $\Omega_g$, which is given by 
\begin{equation}
\Omega_g = \frac{|\bar q| B}{mc}.
\label{eq:Omegag}
\end{equation}
Here $\bar q$ is the time-averaged grain charge, $B$ the magnetic field strength, and $c$ the speed of light.  For the sizes where these timescales are comparable, nearly all grains are either singly negatively charged or neutral (see Table 2; we  note that there are essentially no multiply negatively charged grains).  For this reason, we consider two charging timescales: $\tau_{ch(-1)}$ is the time for a grain of charge -1 to become neutral and $\tau_{ch(0)}$ is the time for a neutral grain to acquire a negative charge. 
\par
The charging timescales, as well as $\tau_s$ and $\Omega_g^{-1}$, are plotted in Figure \ref{fig:timescales} as a function of grain size.   We calculated $\tau_{ch(0)}$ and $\tau_{ch(-1)}$ as a function of grain size using Equations (3.1) - (3.5) from \citet{Draine87}.  These equations ignore the contribution of UV photons from H$_2$ fluorescence due to CR impacts, but this is not important to the charging of the grain sizes where the charging timescale is comparable to the other dynamical timescales.  The top panel is made for the core center, and the bottom is for the envelope.  
We see that for sizes greater than about 10 nm, in both environments, the two charging timescales are shorter than both $\tau_s$ and $\Omega_g^{-1}$.  For that reason, for the calculations in this paper, we use the mean grain charge when calculating the grain dynamics.  This approximation should somewhat {overestimate} the depletion timescale for grains less than 10 nm as in reality there may be significant collision velocities between grains of the same size, but different charge states.  We also performed a simulation in which we considered the opposite limit where we calculate the grain dynamics assuming that an individual grain's charge remains fixed, and found  that the smallest grains were indeed depleted even faster, and that there was little difference for the larger grains.
\par
To estimate the grain velocities, we assume the following model.  Consider a coordinate system in which $\hat r$ points radially outward, and the magnetic field has no radial component.  The ions are assumed to be at rest with respect to the cloud center.  Then, the equations of motion for a dust grain with mean charge $\bar q$ are given by 
\begin{align} 
& \frac{dv_r}{dt} = -g +\Omega_g v_\perp - \frac{v_r + v_{\rm in}}{\tau_s}, 
\nonumber\\
& \frac{dv_\perp}{dt} = -\Omega_g v_r - \frac{v_\perp}{\tau_s},
\end{align}
where $v_\perp$ is the velocity in the direction parallel to $\hat {\bf B} \times \hat {\bf r}$.  In equilibrium, these equations can be solved to yield 
\begin{equation}
 v_r = \frac{-g_{\rm eff} \tau_s}{1+\Omega_g^2 \tau_s^2}, \quad v_\perp = \frac{g_{\rm eff} \Omega_g \tau_s^2 }{1 + \Omega_g^2\tau_s^2} \equiv -v_r \Omega_g \tau_s,
\label{equilibriumSpeeds}
\end{equation}
where $g_{\rm eff} = g + v_{\rm in}/\tau_s$ is the effective gravity.  We note that for the grain sizes considered in this work, $v_{\rm in}/\tau_s \gg g$ for $c_d \leq 1$.  The collision velocity due to these drift effects is 
\begin{equation}
\Delta V_{\rm 12, drift} = \sqrt{(v_{r1} - v_{r2})^2 + (v_{\perp 1} - v_{\perp 2})^2}.
\end{equation}

Small dust grains (those for which $\Omega_g^{-1} \ll \tau_s$), are relatively well coupled to the magnetic field so they move with respect to the gas at velocity $v_{\rm in}$ in the $\hat r$ direction.  Large grains ($\Omega_g^{-1} \gg \tau_s$) do not feel the magnetic field, and drift in the $-\hat r$ direction at speed $g \tau_s$.  The transition between these two regimes occurs near size $a_c$, defined as the size where $\Omega_g \tau_s = 1$.  Using the expressions for $\Omega_g$ and $\tau_s$, we find 
\begin{equation}
a_c = \sqrt{\frac{3|\bar q|B}{4\pi v_{\rm th}^*c\rho_g}}.
\label{acEq}
\end{equation}
We observe that the RHS of Equation \eqref{acEq} also depends  on $a_c$ through $\bar q$.  The mass of the grain does not enter Equation \eqref{acEq}, so the expression for $a_c$ is valid for very porous grains as well.  To estimate $B(r)$ we assume that the core is in hydrostatic equilibrium with the pressure support provided by the magnetic field.  Thus ,$B(r)$ solves the equation
 \begin{equation}
 \frac{d}{dr} \left(\frac{B^2}{8\pi}\right) = -g\rho_g,
 \end{equation}
 where we assume that $B$ outside the cloud is negligible, and $g(r)$ is calculated from the density distribution $\rho_g(r)$ determined by Equation \eqref{TafallaDensity}.  This gives $B$ of 395 $\mu$G in the core center and 45 $\mu$G in the envelope, corresponding to $a_c$ of 34 and 88 nm in the two environments, respectively.  \citet{Nakamura19} measured a line-of-sight magnetic field of 117 $\pm 21$ $\mu$G at an estimated ${\rm H_2}$ density of $3 \times 10^4$ cm$^{-3}$ in the cloud TMC-1.  Multiplying by $\sqrt{3}$ to estimate the total magnetic field gives 203 $\mu G$.  Our model gives 81 $\mu$G at this density.

\subsection{Collision velocity from all mechanisms}
We approximate the collision velocity between two dust particles as
\begin{equation}
\Delta V_{12} = \sqrt{\langle \Delta V_{\rm 12, turb}^2 \rangle + \langle \Delta V_{\rm 12, BM}^2 \rangle + \Delta V_{\rm 12, drift}^2}.
\label{generalCollisionVelocityEquation}
\end{equation}
There are two issues with this expression.  First, it does not necessarily accurately represent the mean collision velocities between bodies in which more than one source of relative motion is important.  Second, it does not account for the fact that if turbulent motions are important, there is  a {distribution} of collision velocities.  Given the large uncertainties on the individual sources of relative motion, both these issues seem rather minor.
\begin{figure}
\centering
\includegraphics[width=1.0\columnwidth]{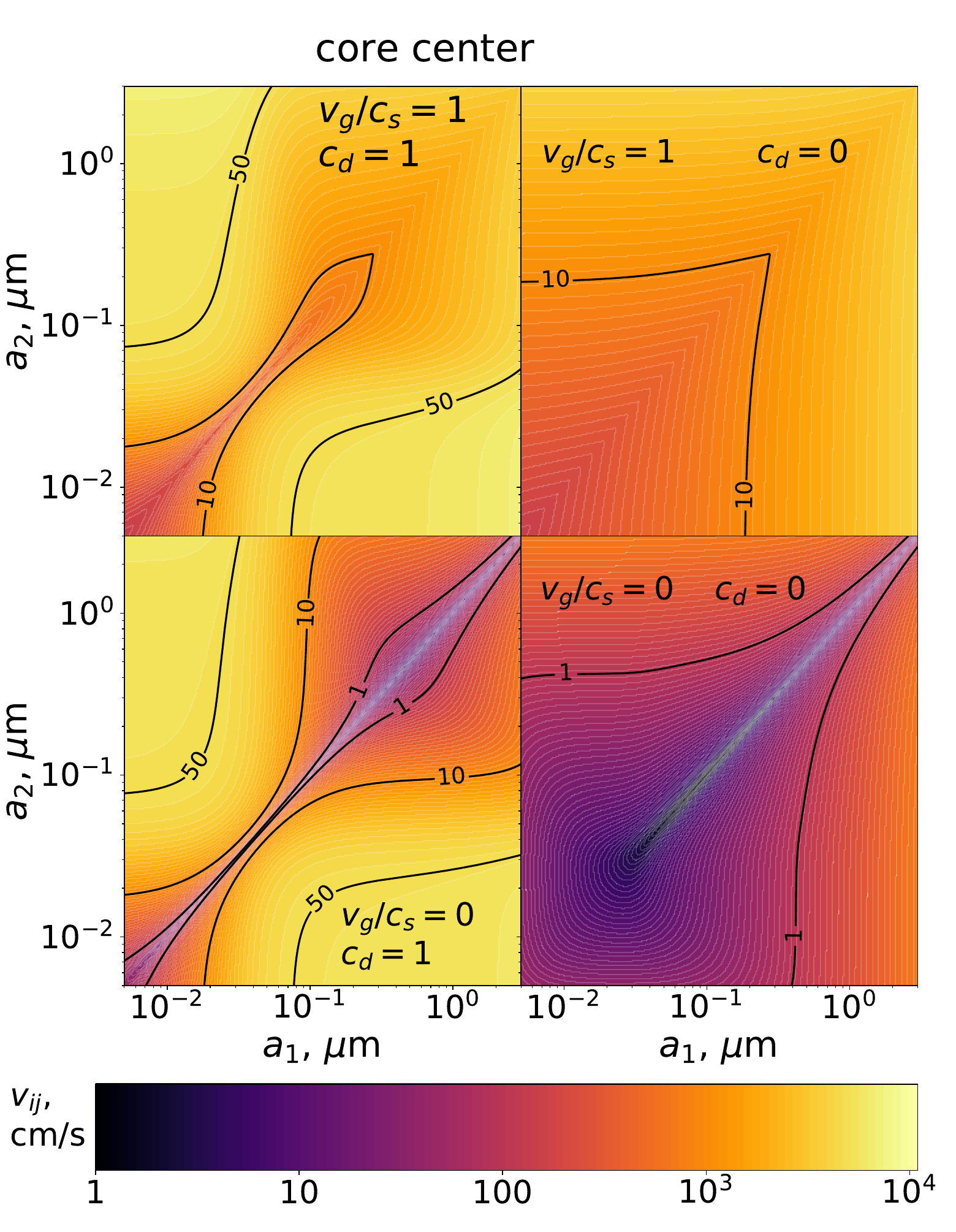}
\caption{Collision velocities calculated from Equation \eqref{generalCollisionVelocityEquation} in the space of the sizes of the collision partners.  The turbulent velocity $v_g$ and the constant $c_d$, which describes the speed of the ion-neutral drift (see Equation \eqref{eq:forceBalance}), are different in each panel, as labeled. The isothermal sound speed  $c_s$ is  given by $\sqrt{k_BT/(2.33 m_p)}$.  The labeled contours are  at 1, 10, and 50 meters per second.}
\label{collisionVelFig}
\end{figure}

\begin{figure}
\centering
\includegraphics[width=1.0\columnwidth]{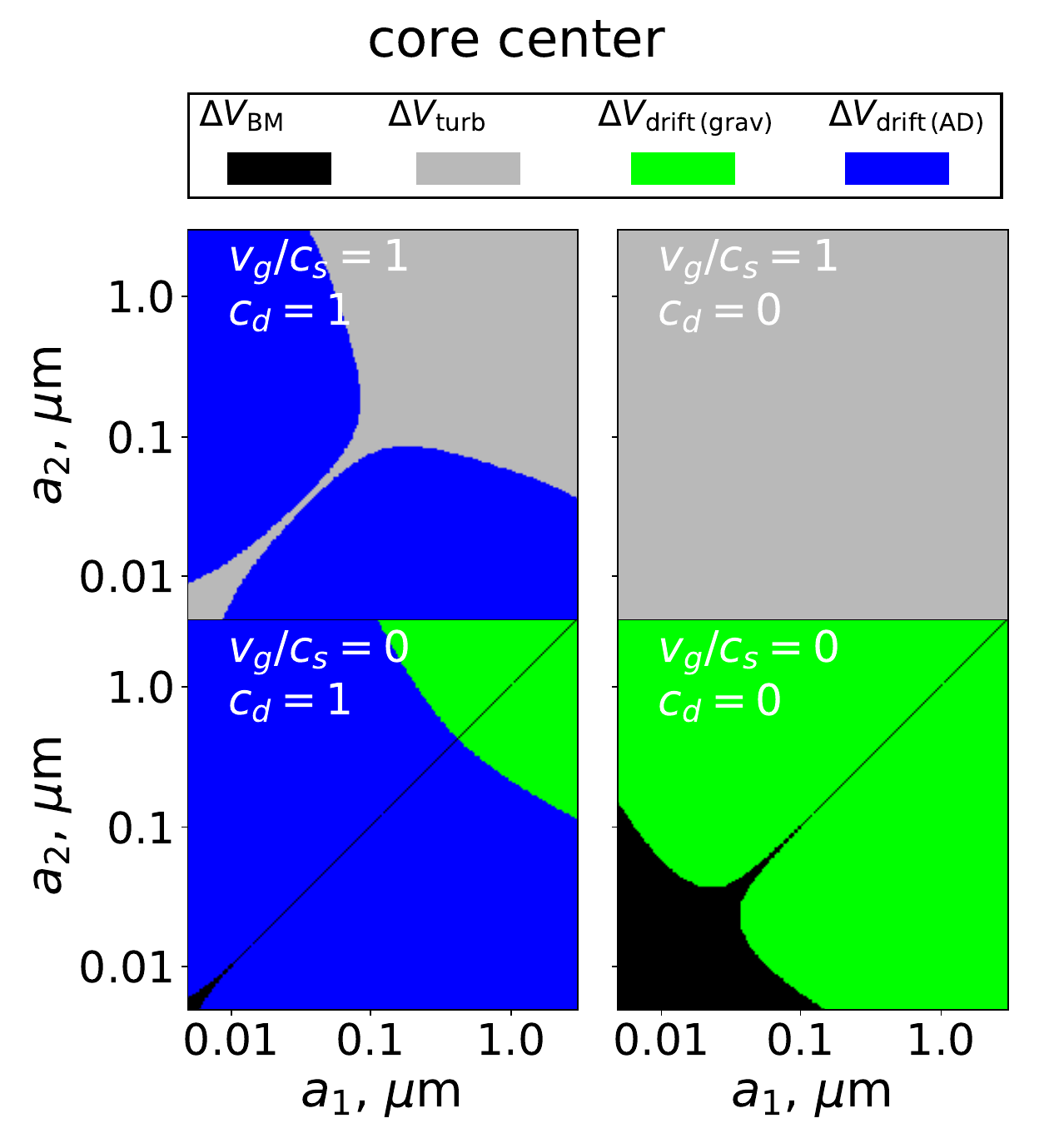}
\caption{Dominant source of collision velocity in the space of the sizes of the collision partners.  The turbulent velocity $v_g$ and the constant $c_d$, which describes the speed of the ion-neutral drift [see Equation \eqref{eq:forceBalance}], are varied between the panels, as labeled.  $v_g$ and $c_d$ in each panel take the same values as in the corresponding panel of Figure \ref{collisionVelFig}.  The mapping between color and dominant velocity source is shown in the legend on top.}
\label{velocityRegimesFig}
\end{figure}
\par
Figure \ref{collisionVelFig} shows the collision velocities as a function of the sizes of the collision partners.  As we are uncertain about the strength of the turbulence and the ambipolar diffusion, the four panels show the four possible cases in which each effect is either present or absent.  We see that collision velocities are generally highest for collisions in which one grain is larger than $a_c$ and one is smaller, or between two large ($\gtrsim 1 \mu$m) grains in the models where turbulence is present. Figure \ref{velocityRegimesFig} shows the dominant source of collision velocity for each set of collision partners. Grain collision velocities are dominated by turbulence and ambipolar diffusion.  Gravitational settling is dominant for a wide range of particle sizes only if there is no turbulence and we ignore ambipolar diffusion.  Brownian motion is typically irrelevant for all except the smallest particles.  Turbulence and Brownian motion are able to excite collision velocities between grains of the same size.  Systematic drifts are not.  For that reason, we see narrow gray  and black lines across the panels, corresponding to same-size collision partners where the systematic drifts have no effect.

\subsection{Collision rates}
In our dust coagulation code the size distribution is discretized in $N_{\rm bins}$ discrete logarithmically spaced size bins between $r_{\rm min}$ and $r_{\rm max}$.  We can then write the number of collisions per unit time between bodies in mass bin $i$ and $j$ in a given thin spherical shell as 
\begin{equation}
\frac{d N_{\rm coll}}{dt} = \frac{N_i N_j }{V} \sum_{q_i, q_j}v_{ij}(q_i, q_j) \sigma_{ij}(q_i, q_j) f_{q_i}f_{q_j}.
\end{equation}
 Here $N_i$ and $N_j$ are the numbers of particles in mass bin $i$ and $j$, respectively, and  $v_{ij}(q_i, q_j)$ is the collision velocity between a particle in mass bin $i$ with charge $q_i$ and a particle in mass bin $j$ with charge $q_j$.  We note  that for our standard assumption that the charging timescales are rapid compared to the other dynamical timescales, $v_{ij}$ is calculated using $\bar q_i$ and $\bar q_j$; however, $\sigma_{ij}$ is still calculated using the instantaneous charge states $q_i$ and $q_j$.  The parameter $f_{q_i}$ is the fraction of particles in mass bin $i$ with charge $q_i$, and $f_{q_j}$ is defined analogously for mass bin $j$;   $V$ is the volume of the shell.  The collision cross section is given by 
  \begin{equation}
 \sigma_{ij}(q_i, q_j) = \pi (a_i + a_j)^2 \max \left \{0,  \, 1-\frac{2q_iq_j}{\mu v_{ij}^2 (a_i + a_j)}\right \},
 \label{eq:Sigma}
 \end{equation}
 where $a_i$ and $a_j$ are the radii of the two dust grains.  We find in Section \ref{sect:zeta} that Coulomb attraction and repulsion play a negligible role in the environments we consider\footnote[1]{Coulomb repulsion can inhibit the coagulation of micron-sized grains in protoplanetary disks, where the expected turbulence is relatively weak \citep{Okuzumi09}.}. This is to be expected given the typical collision velocities.  As can be seen from Equation \eqref{eq:Sigma}, the critical parameter determining the Coulomb effect is $2q_iq_j/\left(\mu v_{ij}^2 (a_i + a_j) \right)$.  For 10 nm grains, this is less than unity for $v > 2$ m s$^{-1}$.  It scales inversely with the fourth power of the sizes of the colliding grains, so it will be smaller for larger grains.  Figure \ref{collisionVelFig} shows that the typical collision velocities are substantially larger than 2 m s$^{-1}$, except in the case with no turbulence or ambipolar diffusion.

\section{Characteristic timescales}
\label{sect:timescales}
In this section we estimate four relevant timescales.  We consider $\tau_{\rm AD}$, the time for small grains to collide with large ones due to ambipolar diffusion; $\tau_{\rm grav}$, the growth timescale of the largest grains due to gravitational settling; $\tau_{\rm turb}$, the growth timescale of the largest grains due to turbulent motions; and  $\tau_{\rm cl}$, the lifetime of the cloud against ambipolar diffusion, which we take as the runtime of our simulations.
\subsection{Cloud lifetime}
\label{sect:lifetime}
We assume that the lifetime of the cloud at scale $r$ is  roughly equal to the ambipolar diffusion time at that scale in the cloud \citep{Tassis04}, i.e., the cloud lifetime $\tau_{\rm cl}$ is given by 
\begin{equation} 
\tau_{\rm cl} = r/v_{\rm in}
\label{eq:tauc}
.\end{equation} 
This works out to be $2.6 \times 10^5$ years in the core center, and $2.3 \times 10^6$ years in the envelope, assuming the standard value for the ionization fraction given in Equation \eqref{eq:ionFrac}.
\subsection{Elimination of small grains due to ambipolar diffusion drift}
\label{sect:tauad}
We can estimate the timescale on which grains coupled to the magnetic field collide with grains coupled to the neutral gas as follows.  First, by setting the mass in dust equal to a fraction $f_d$ of the gas mass, and assuming an MRN distribution \citep{Mathis77}  between sizes $a_{\rm min}$ and $a_{\rm max}$ ($dn/da \propto a^{-3.5}$), we can write the differential density of dust grains with radius $a$ as 
\begin{equation}
n_d(a) = \frac{3\rho_gf_da^{-7/2}}{8\pi\rho_d \sqrt{a_{\rm max}}},
\end{equation}
where $f_d = 0.01$ is the gas-to-dust mass ratio.  Then, we consider a small grain of size $a_s < a_c$.  We  approximate its collision velocity with another larger grain of size $a_l$ as 0 for $a_l < a_c$ and $v_{\rm in}$ for $a_l > a_c$.  We  approximate the collision  cross section as $\pi a_l^2$.  Then the expected time for the smaller grain to experience a collision with a larger one is given by 
\begin{equation}
\tau_{\rm AD}^{-1} = v_{\rm in} \int_{a_c}^{a_{\rm max}} \pi a_l^2 n_d(a_l) da_l.
\end{equation}
This gives 
\begin{equation}
\tau_{\rm AD} = \frac{4 \rho_d \sqrt{a_{\rm max} a_c}}{3\rho_gf_dv_{\rm in}}.
\label{eq:tausl}
\end{equation}
For the parameters used here, we find $\tau_{\rm AD} = 1.1 \times 10^4$ years in the core center, and $2.4 \times 10^6$ years in the envelope.  \subsection{Coagulation of large grains due to gravitational settling}
We can also estimate the time required for the upper end of the mass distribution to grow due to collision velocities arising from gravitational settling.  We assume in this case that all grains are large enough that we may ignore the contribution of the magnetic field.  In this case, the appropriate limit of Equation \eqref{equilibriumSpeeds} ($\Omega_g \rightarrow 0$) yields a collision speed between two particles of sizes $a_s$ and $a_l$ of 
\begin{equation}
v_{\rm grav} = \rho_d g \left(a_l - a_s\right)/(v_{\rm th}^* \rho_g). 
\label{eq:vgrav}
\end{equation}
\par
Then, assuming the collision cross section to be $\pi a_l^2$, dropping the term $a_s$ in Equation \eqref{eq:vgrav}, and taking for simplicity the lower end of the mass spectrum to be $0$, we can calculate a growth timescale for bodies initially at the top end of the mass distribution:
\begin{equation}
\tau_{\rm grav} = M/\dot M = \frac{4v_{\rm th}^*}{3gf_d}.
\label{eq:taugrav}
\end{equation}
This expression is equal to $1.9 \times 10^6$ years in the core center and $8.4 \times 10^6$ years in the envelope.
\subsection{Coagulation of large grains due to turbulence}
In the limit of small Stokes numbers,  Eqs. 6, 7, and 10 from \citet{Ormel07} show that the relative velocity between two grains of the same size is
\begin{equation}
\Delta V_{\rm 12, drift}^2 = v_g\sqrt{\psi(\epsilon) St_l}, 
\label{eq:smallStokes}
\end{equation}
where $St_l$ is the Stokes parameter of the larger grain, $\epsilon$ is the ratio of the size of the smaller grain to that  of the larger, and $\psi(\epsilon)$ varies between 1.95 for $\epsilon = 1$ and 2.95 for $\epsilon = 0$.  To calculate the growth timescale we assume that growth is dominated by collisions of similar-sized grains.  We therefore assume a cross section of $4 \pi a_l^2$, and calculate the velocities using Equation  \eqref{eq:smallStokes} with $\psi = 1.95$.  These approximations yield a timescale for growth of the largest grains from collisions arising from turbulence:
\begin{equation}
\tau_{\rm turb}(a_l) = \left(216 \psi^2 \pi^2 \right)^{-1/4}\sqrt{\frac{\rho_d v_{\rm th} a_lr}{f_d^{*2} \rho_g v_g^3}}
\label{eq:tauturb}
.\end{equation}
Here $r$ enters   through the assumption that it is equal to the turbulence injection scale.  Assuming $v_g = c_s$, where $c_s = \sqrt{k_BT/(2.33 m_p)}$, this expression gives $9.4 \times 10^4$ years in the core center, and $2.0 \times 10^6$ years in the envelope.  Given our belief that turbulent velocities near the core center are subsonic, the core center timescale represents a lower limit to the probable actual coagulation timescale.

\par
Using Equation \eqref{eq:tauturb} in conjunction with the fact that $a/\dot a = 3 m/\dot m$, we can solve for $a_l(t)$, assuming that $a_l(t_0) = a_{\rm max}$:
\begin{equation}
a_l(t) = a_{\rm max} \left[1 + \frac{t}{6 \tau_{\rm turb}(a_{\rm max})} \right]^2
\label{eq:peak}
.\end{equation}
We show in Section \ref{sect:zeta} that this is a reasonable approximation, but yields a slight overestimate.

 \section{Results}
 \label{sect:results}
 \subsection{Core center}
\begin{figure}
\centering
\includegraphics[width=1.0\columnwidth]{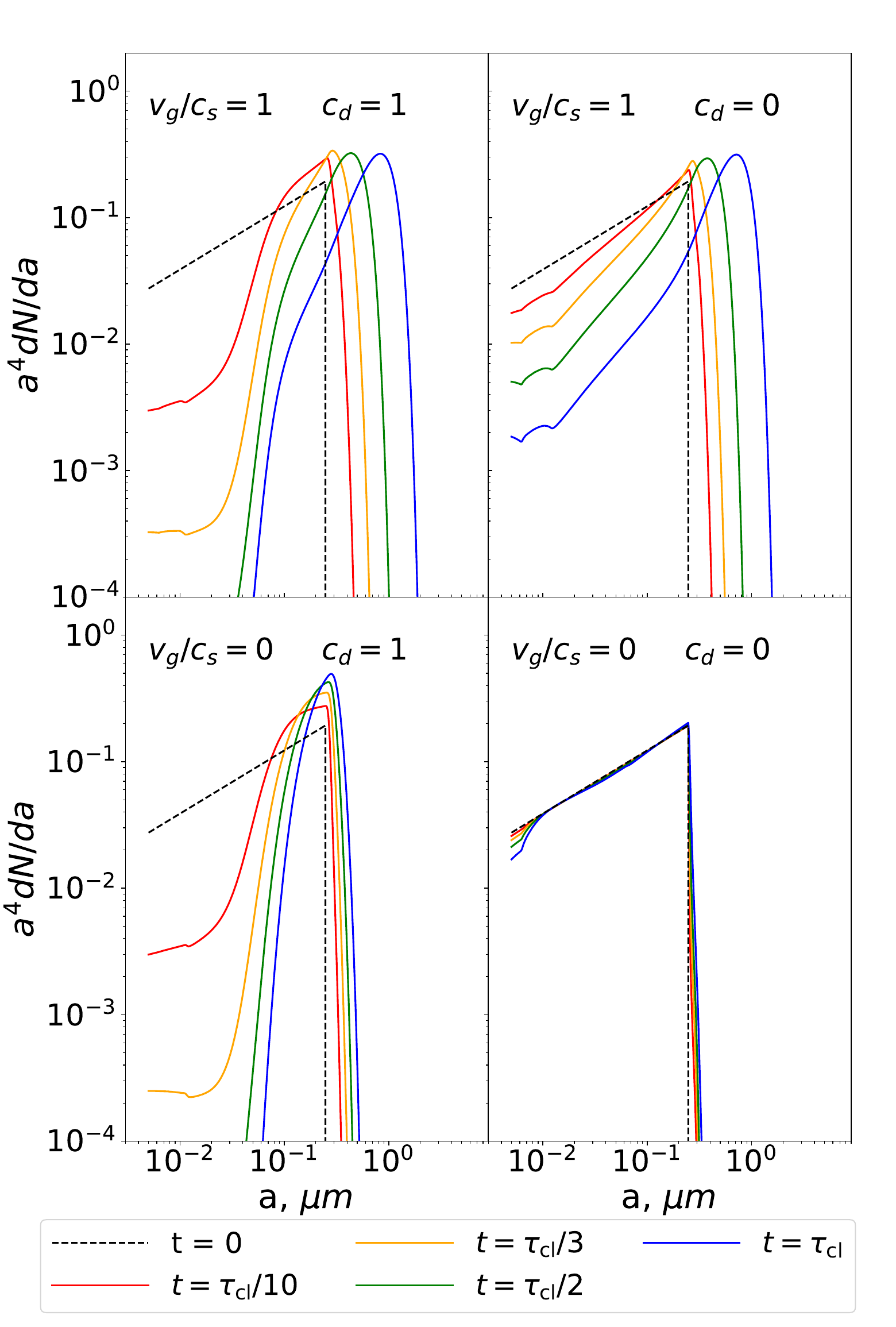}
\caption{Evolution of the grain-size distribution in the core center for different values of the parameters $v_g$ and $c_d$, as labeled in the panels.  The different color curves correspond to different amounts of time the simulation was run for in terms of the cloud lifetime $\tau_{\rm cl}$ [see Equation \eqref{eq:tauc}].  Here $\tau_{\rm cl} = 2.6 \times 10^5$ years.  In each case the initial distribution is an MRN distribution from 5 to 250 nm, which is shown as the black dashed line. }
\label{fig:core}
\end{figure}

We consider here the coagulation of dust at radius $r_0$ from the center.  We consider four models with different values of $v_g$ and $c_d$, the parameters determining the strength of the turbulence, and the ion-neutral drift velocity.  The results are shown in Figure \ref{fig:core}.  The values of $v_g$ and $c_d$ in each panel are equal to those in the corresponding panel of Figure \ref{collisionVelFig}.  Different color curves correspond to the size distribution at the time shown in the legend.  The dashed black curve is the MRN distribution, used as the initial condition. 
\par
The simulation was run for time $\tau_{\rm cl}$ (see Section \ref{sect:lifetime}), equal in this case to $2.6 \times 10^5$ years. In all cases, for times shorter than $\tau_{\rm cl}/3$ years, there is very little evolution of the upper end of the size distribution.  The evolution of the high end of the size distribution depends most on the value of $v_g$, which makes sense as we see in Figure \ref{velocityRegimesFig} that collisions between bodies with sizes around 1 $\mu$m depend primarily on the strength of the turbulence.  In Section \ref{sect:timescales} we derived growth timescales of $9.4 \times 10^4$ years and $1.9 \times 10^6$ years (40\% and 730\% of $\tau_{\rm cl}$) for growth of the upper end of the size distribution due to turbulence and gravitational settling, respectively, assuming sonic turbulence.  This is consistent with the growth of the peak of the distribution to about a micron for the simulations with sonic turbulence, and only minor changes to the upper scale of the distribution for all simulations with subsonic turbulence.  It should be noted from Equation \eqref{eq:tauturb} that $\tau_{\rm turb} \propto v_g^{-3/2}$.  Examination of the bottom panels of Figure \ref{fig:core} shows that in the absence of turbulence, ambipolar diffusion can lead to a minor change in the upper edge of the size distribution, but cannot produce  micron-sized grains.
\par
We find when $c_d = 1$ that there is a rapid depletion on $\sim 10^4$ year timescales of grains less than $a_c$ (defined by Equation \ref{acEq}), consistent with the estimate in Equation \eqref{eq:tausl}.  The bottom right panel of Figure \ref{fig:core}, shows that Brownian motion has   a minor effect, and only for the very smallest grains.

\subsection{Envelope}
\begin{figure}
\centering
\includegraphics[width=1.0\columnwidth]{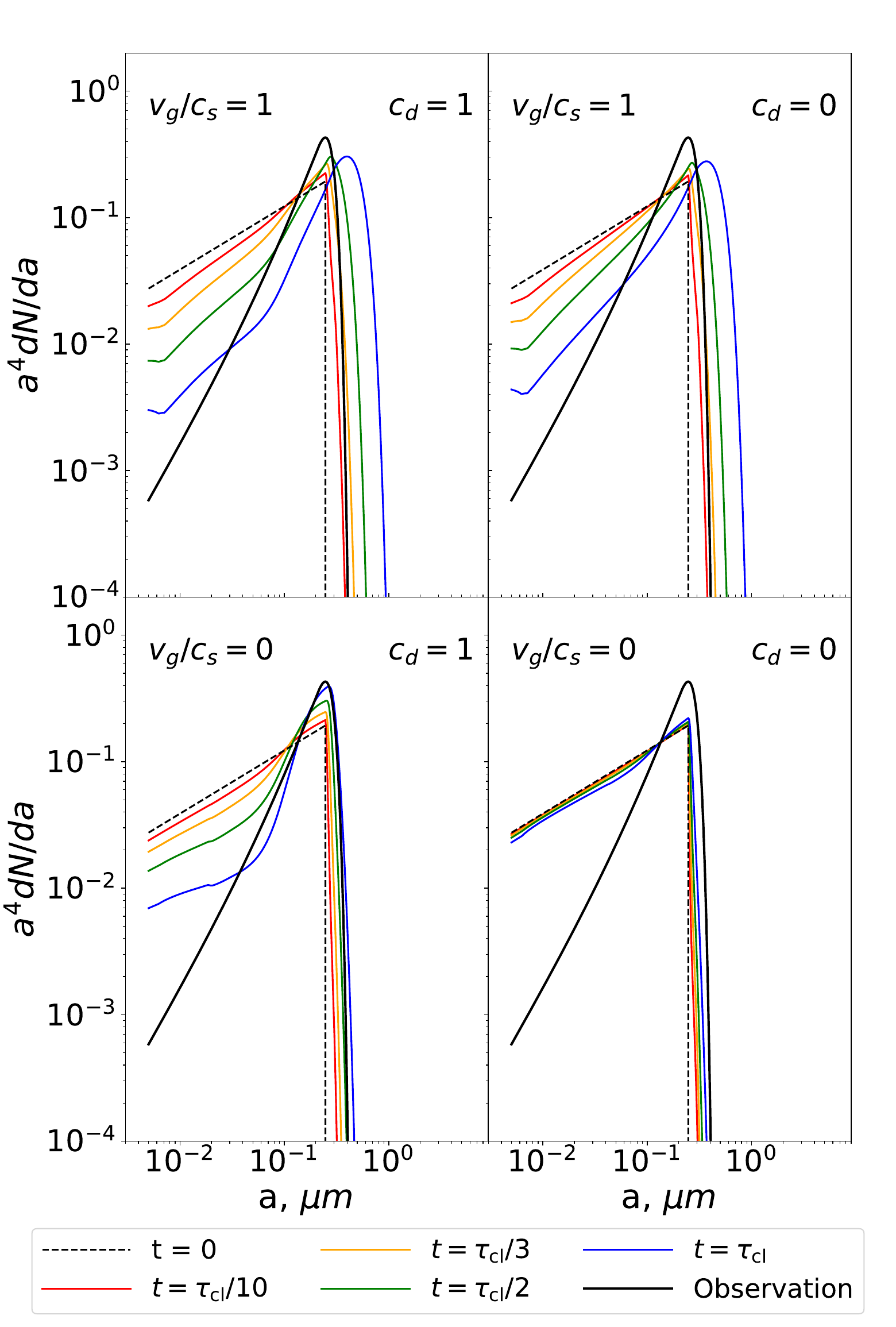}
\caption{Same as Figure 4, but for the envelope.  The colored curves show the size distribution at different times.  In this case $\tau_{\rm cl} = 2.3 \times 10^6$ years.  The black solid line shows the curve inferred in \citet{Weingartner01} for dense regions.}
\label{fig:envelope}
\end{figure}

We also consider coagulation in a lower density environment ($n_{H_2} = 10^4$ cm$^{-3}$).  This corresponds to a location at $r = 20,\!800$ AU in the model of \citet{Tafalla02}.  Figure \ref{fig:envelope} is analogous to Figure \ref{fig:core}, but made for the envelope.  As discussed in Section \ref{sect:timescales}, the cloud lifetime $\tau_{\rm cl}$ is higher by nearly a factor of 10 in this environment.  This does not, however, lead to more growth because $\tau_{\rm turb}$ is higher by a factor of $\sim 20$, and $\tau_{\rm AD}$ higher by a factor of 200. The value of  $\tau_{\rm grav}$ is higher by only a factor of $4$, but is still longer than $\tau_{\rm cl}$.  For this reason, there is somewhat less growth of the top of the dust distribution, and significantly less complete removal of the small grains.
\par
 The solid black curve is the curve derived by \citet{Weingartner01} for dense regions (their ``case A'' model with $R_v = 5.5$ and $10^5 b_c = 3.0$).  It is not clear   which region of the cloud these observations are expected to best correspond to.  We see none of the models are able to reproduce the absence of the smallest grains from \citet{Weingartner01}.  The ones which do the best are those with high values of $c_d$.  There is also only minimal growth of the large end of the spectrum, consistent with the model from \citet{Weingartner01}, but suggesting that the coreshine observations from \citet{Pagani10} must have come from dust in regions denser than $10^4$ cm$^{-3}$ or that our estimate of the cloud lifetime is too short (see next subsection).
\subsection{Effect of CR ionization rate}
\label{sect:zeta}
Cosmic rays can affect the coagulation process in two ways.  First, they are the primary cause of gas ionization, and therefore determine the ambipolar diffusion velocity $v_{\rm in}$ defined in Equation \eqref{eq:forceBalance}.  Since the ionization fraction is assumed to be proportional to $\sqrt{\zeta}$, this means that $v_{\rm in} \propto 1/\sqrt{\zeta}$.  As the cloud lifetime $\tau_{\rm cl}$ is proportional to $1/v_{\rm in}$ (see Section \ref{sect:lifetime}), there is more time for coagulation to occur at higher $\zeta$.  We note, however, that $\tau_{\rm AD}$ also scales as $1/v_{\rm in}$ (see Section \ref{sect:tauad}).  Hence, to the zeroth order, the degree to which small grains are eliminated in the time $\tau_{\rm cl}$ due to ambipolar diffusion drift is independent of $\zeta$.
\par
Second, as discussed in Section \ref{sect:grainCharge}, CRs also excite electronic states of ${\rm H_2}$ that decay, producing UV photons that lead to more positively charged dust grains \citep{Prasad83, Gredel89, Ivlev15}.  

 \begin{figure}
\centering
\includegraphics[width=1.0\columnwidth]{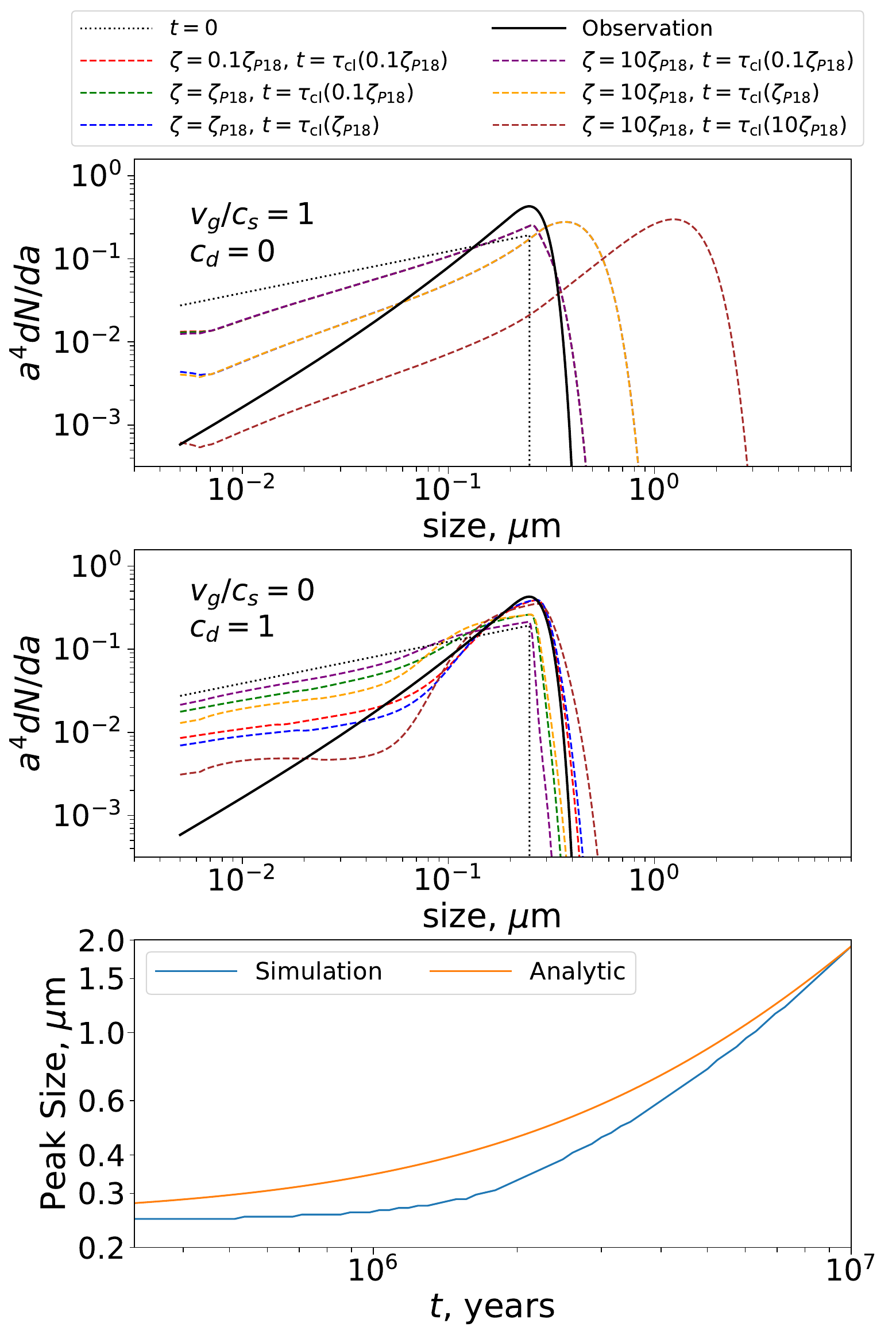}
\caption{Top panel: Size distribution in the envelope for the case with sonic turbulence and $c_d = 0$.  The different curves correspond to different amounts of time and different CR rates (see legend at top).  Time  is expressed in terms of $\tau_{\rm cl}$ {for each different CR ionization rate.} $\tau_{\rm cl}(0.1 \zeta_{P18})$, $\tau_{\rm cl}(\zeta_{P18})$, and $\tau_{\rm cl}(10 \zeta_{P18})$ are the cloud lifetimes, evaluated assuming CR ionization rates of $0.1 \zeta_{\rm P18}$, $\zeta_{\rm P18}$, and $10 \zeta_{\rm P18}$, respectively, where $\zeta_{\rm P18}$ is the ionization rate calculated from \citet{Padovani18}.  In the case of $c_d = 0$ (top panel)   the curves at different times are nearly independent of the ionization rate and the curves overlap nearly perfectly (i.e., only one curve  is visible for each time).  Middle panel: Without turbulence, but assuming $c_d = 1$.  Bottom panel:   Shown in blue is the peak of the distribution as a function of time (assuming $\zeta = \zeta_{\rm P18}$, $c_d = 0$, and $v_g = c_s$), and in orange  the approximation given by Equation \eqref{eq:peak}.}
\label{fig:growthPlot}
\end{figure}

To examine the effect of changing $\zeta$ on the evolution of the size distribution, we ran the coagulation simulation for different CR ionization rates.  These runs were done for the envelope, where the CRs have more effect on the dust charge distribution.  The results are shown in the top two panels of Figure \ref{fig:growthPlot}.  Different curves correspond to different assumed ionization rates and simulation run times.  The simulation run time is expressed in units of the cloud lifetime $\tau_{\rm cl}$, given in Equation \eqref{eq:tauc}.  Because $v_{\rm in} \sim \zeta^{-1/2}$,  the cloud lifetime is proportional to $\zeta^{\frac{1}{2}}$.
\par
The top panel assumes sonic turbulence and $c_d = 0$.  In this case we see that the coagulation process is essentially unaffected by $\zeta$, except that a higher $\zeta$ means a long cloud lifetime in our model.  This shows that the issue of Coulomb focussing--repulsion is completely unimportant, making only very minor changes to the size distributions at the smallest sizes.
\par
The middle panel assumes no turbulence, and $c_d = 1$.  In this case, naively we would expect little difference in the degree of coagulation within the cloud lifetime since the collision velocities are dominated by ambipolar diffusion, so the cloud lifetime is inversely proportional to the scale for the collision velocities.  Instead, we see that over the lifetime of the cloud, the high $\zeta$ model still gets rid of more of the small grains because a higher CR rate decreases the mean grain charge for grains with sizes between 10 and 100 nm (see Table 2), and hence $a_c$ decreases too (see Equation \eqref{acEq}).  Equations \eqref{eq:tauc} and \eqref{eq:tausl} demonstrate that $\tau_{\rm AD}/\tau_{\rm cl} \propto \sqrt{a_c}$, and therefore also decreases with increasing $\zeta$. 
\par
The ability to grow micron-sized grains depends sensitively on the cloud lifetime $\tau_{\rm cl}$.  Equation \eqref{eq:peak} shows that the peak size roughly scales as the square of the elapsed time for large times if growth is dominated by turbulence.  The peak of the grain mass distribution (peak of $m^2 dN/dm$) is shown by the blue curve in the bottom panel of Figure \ref{fig:growthPlot}, and the approximation given by Equation \eqref{eq:peak} is shown in orange.  The orange curve gives a slight overestimate for two reasons.  First, there is initially no movement of the peak due to the sharp cutoff of the original size distribution.  Additionally, Equation \eqref{eq:peak} assumes that all collisions are between objects at the peak size, so the cross section is slightly overestimated.  Nonetheless, it is clear that were the cloud to survive a few times longer than in our model, the peak size would be substantially larger.

 \section{Additional effects}

 \subsection{Evolution in  the  core center and the envelope}
In reality, material goes through a low-density phase better described by our envelope model before becoming part of the core center.  To model this we took the output of our envelope simulations as the initial input to the core center simulations.  We found, however, that the preprocessing in the envelope made very little difference to the resulting distribution after coagulation within the core center.  This makes sense as the evolution is qualitatively similar, but the lifetime of the core center is longer in units of the local collision timescale than the lifetime in the envelope in units of the envelope collision timescale.  
\subsection{Contribution of small grains to ambipolar diffusion velocity}
\label{sect:smallGrainAmbi}
In Equation \eqref{eq:forceBalance} we calculate the ambipolar drift speed, assuming that neutrals are slowed down only through collisions with ions.  Neutrals must to some extent be slowed by collisions with small magnetically coupled dust grains as well.  To calculate this, we can add the term
\begin{equation}
 \int_{a_{\rm min}}^{a_{\rm max}} \pi a^2 n(a) v_{\rm dn}(a) \left[m_{\rm H_2} n_{\rm H_2} \langle v_{\rm H_2}\rangle +  m_{\rm He} n_{\rm He} \langle v_{\rm He} \rangle \right]da
 \label{eq:dustCoupling}
 \end{equation}
  to the right-hand side of Equation \eqref{eq:forceBalance}, where $v_{\rm dn}(a)$ is the speed of a dust grain with radius $a$ with respect to the neutral gas.  This is not a significant effect in the envelope.  Even if all the dust grains were completely coupled to the magnetic field, their addition to the RHS of equation \eqref{eq:forceBalance} would be smaller than that of the ions by more than a factor of 5, assuming an MRN distribution.
\par
On the other hand, this additional coupling may be significant in the core center where the ion density is significantly lower.  Let us denote the ratio of the expression in Equation \eqref{eq:dustCoupling} to the right-hand side of Equation \eqref{eq:forceBalance} as $\chi$.  An upper bound on $\chi$ can be found by assuming an MRN distribution, and that the grains are fully coupled, so $v_{\rm dn} = v_{\rm in}$.  In this case $\chi = 5$.  In reality for an MRN distribution, $\chi = 3.5$ because the larger grains are uncoupled.  Nevertheless, this is still a significant correction.  However, the problem resolves itself rapidly.  After a time $\tau_{\rm cl}/10$, the simulation with no turbulence, but $c_d = 1$ has removed a sufficient number of the small grains so that $\chi = 0.4$.  Of course the removal is somewhat slower (by a factor of $1 + \chi$) in reality when the small magnetically coupled grains are still present, but it seems reasonable that for most of the evolution time of the cloud the magnetically coupled grain population would be small enough that we are justified in ignoring this correction (or lumping it together with the fudge factor $c_d$).  Additionally, Figure \ref{fig:envelope} shows that there is a substantial reduction in the number of small grains even in the envelope stage, so by the time grains reach the core center, the density of magnetically coupled grains should be reduced by a factor of several.

\subsection{Icy mantles and non-compact grains}
\label{sect:mantles}
In the interests of simplicity, the preceding analysis leaves  out two important effects that may enhance dust coagulation.  The first is that at densities higher than $\sim 10^5$ cm$^3$, nearly all of the carbon and oxygen in the gas phase freezes out onto the dust grains \citep{Caselli99}.  This results in a substantial increase in the grain volume.  To estimate this we assume that all the carbon and oxygen in the ISM condenses onto the surface of the grains.  We assume that the ratio of carbon and oxygen to hydrogen is the same as in the Sun.  Using the abundances given in \citet{Asplund09}, and assuming a density of 1g per cm$^3$ for this mantle material, we find that the total volume of the grains is increased by a factor of $V_{\rm rel} = 2.8$.  This factor is substantially uncertain; for reference, \citet{Ossenkopf93} considers two cases in which the mantle component has 0.5 times and 4.5 times the volume of the refractory cores.  
\par
Additionally, realistic grains are not perfect compact spheres, and therefore have a higher projected area-to-mass ratio.  \citet{Shen08} consider model grains created through ballistic aggregation followed by compaction.  They find, depending on the model, a ratio of angle-averaged projected area to the area of a solid sphere of the same material between 1.4 and 4.  In this section we take this to be a factor of 2.  To account for this, we increase our collision cross sections, and decrease our stopping time by a factor of 2.
\par
 \begin{figure}
\centering
\includegraphics[width=1.0\columnwidth]{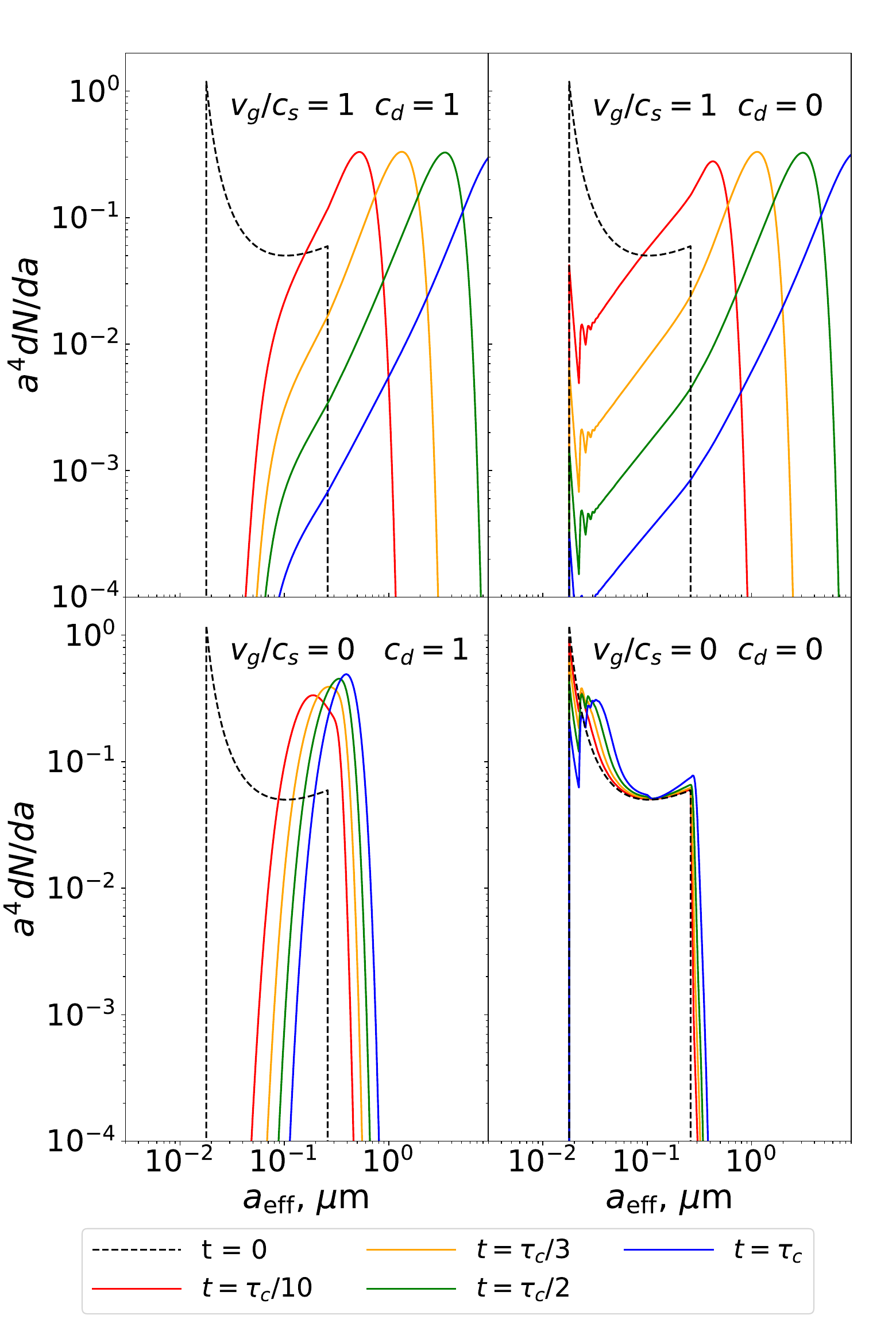}
\caption{Same as Figure \ref{fig:core}, but assuming that grains are covered with icy mantles, and that they are non-spherical and/or porous (which increases their projected surface area by a factor of 2).  As in Figure \ref{fig:core}, the  panels show different degrees of turbulence and ambipolar diffusion, parameterized by $v_g$ and $c_d$, as labeled.   The sharp peak around 20 nm is the result of the assumption that mantles of equal thickness were applied to all the grains in the MRN distribution.  Different ways of assigning the mantle volume do not appreciably alter the results at later times.  The wiggles seen in some of the curves near 20 nm are a simulation effect to do with the sharp cutoff of the initial mass distribution and the finite bin width.  }
\label{fig:mantle12}
\end{figure} 
Figure \ref{fig:mantle12}  is analogous to Figure \ref{fig:core}, but as discussed above   includes the icy mantles and considers the effect of non-spherical and/or porous grains on the dynamics and collision rates.  We use a couple of approximations in our treatment of the mantles.  At $t = 0$, we assume that the refractory cores have an MRN size distribution.  In addition, each grain gets a mantle of a constant thickness $a_{\rm mantle} = 13$nm, which is chosen so that the ratio of the mantle volume of the whole grain population to the core volume of the whole grain population is $V_{\rm rel} = 2.8$.  This leads to most of the volume being in grains of size between 1 and 2 times the mantle thickness, as can be seen in Figure \ref{fig:mantle12}.  Constant mantle thickness is the result we would get if there were no grain-size dependence to the freeze-out process.  In reality, the smaller grains may have their mantles evaporate due to transient CR heating \citep{Leger85, Zhao18b}.  We find that the initial way the mantle material is distributed does not matter.  At time $\tau_{\rm cl}/10$, all memory of the spike in the size distribution at the smallest sizes has been removed for the cases with $c_d = 1$, and even with just turbulence,  the small sizes no longer dominate the grain volume.  We verified that the way the mantle volume is initially distributed has a negligible effect on the size distribution at later times.
\par
In determining the grain dynamics, we assume that for grains of size between $a_{\rm min}$ and $a_{\rm max}$,
\begin{equation}
\rho = \rho_{\rm core}\left(1-\frac{a_{\rm mantle}}{a}\right)^3 + \rho_{\rm mantle} \left[1 - \left(1 - \frac{a_{\rm mantle}}{a}\right)^3\right].
\end{equation}
For sizes $a > a_{\rm max}$ the density smoothly approaches the mean density given by 
\begin{equation}
\bar \rho = \frac{\rho_{\rm core} + V_{\rm rel} \rho_{\rm mantle}}{1+V_{\rm rel}}.
\end{equation}
Comparing Figure \ref{fig:mantle12} with Figure \ref{fig:core}, it is apparent that growth to above one micron within the estimated core lifetime is only possible if we consider the icy mantles and non-spherical and/or porous grain shapes.  With these factors taken into consideration, we find that growth to about ten microns occurs in our models with sonic turbulence.  With just ambipolar diffusion, the differences between Figure \ref{fig:mantle12} and Figure \ref{fig:core} are less extreme.   The evolution proceeds faster with the icy mantles, but the end result in both cases is a narrow distribution centered near the top of the initial distribution.
\subsection{Grain fragmentation}
A significant uncertainty which we have not considered is the possibility of collisional outcomes other than perfect sticking.  There have been significant efforts both experimentally and theoretically to characterize the outcome of grain-grain collisions, but it remains extremely uncertain (see \citealt{Blum08} for a good review).  It is likely that silicate grains would fragment given our collision speeds of up to 50 m s$^{-1}$.  However, the situation may be different if the grains are coated with icy mantles.  \citet{Wada13} performed simulations of particle aggregates with various compositions, and found that icy aggregates grew in collisions with velocities below 80 $\left(a/0.1\mu {\rm m}\right)^{-5/6}$ m s$^{-1}$ (assuming that the aggregates were composed of spherical monomers with radius $0.1\, \mu$m).  For bare silicate grains, the critical velocity was an order of magnitude lower.  \citet{Kimura15} found a critical velocity of 4 m/s for $0.1\, \mu$m bare silicate spheres, but argued that coating the grains in water ice does not help very much.  On the other hand, they also studied a model in which the grains are coated in complex organic mantles, and found a critical sticking velocity of 66 m/s.  \citet{Ormel09}, using a model of turbulent collision velocities similar to ours, concluded that ice-coated grains reach $\sim 100$ $\mu$m before fragmentation prevents further growth, whereas for pure silicate grains this occurs at around a micron.

\par

 On the experimental side, \citet{Gundlach15} find a critical velocity for 1.5 $\mu$m ${\rm H_2O}$ spheres of 9.6 m s$^{-1}$.  Smaller particles are thought to be more likely to stick.  Extrapolating the experimental result from \citet{Gundlach15} down to 0.1 $\mu$m spheres (using the $\propto a^{-5/6}$ scaling) would suggest a critical sticking velocity of 90 m s$^{-1}$.  Not all refractory materials, however, have such a high sticking threshold.  \citet{Musiolik16} found a critical velocity of 0.04 m s$^{-1}$ for 100 $\mu$m CO$_2$ grains, and noted that it would be nearly an order of magnitude lower if these grains were scaled to 1.5 $\mu$m.

\par

  Given such a wide range of available results in the literature, and noting that we are primarily concerned with grains $\leq 0.1\, \mu$m, we neglect the fragmentation effect in the present paper and assume a perfect sticking of colliding grains.
\section{Some implications of this work}
 \subsection{Gas temperature}
 \label{sect:temperatures}
  \citet{Ivlev19} has shown that in the core center the gas temperature is sensitive to the dust grain-size distribution.  At these densities, the gas heating is dominated by CRs, which heat the gas at a rate proportional to the ionization rate, and is  only weakly dependent on other parameters \citep{Glassgold12}.  The gas cools primarily due to collisions with dust grains.  If the dust has coagulated into larger bodies, then there is less surface area for the gas to collide with, so the cooling is reduced.  In addition to grain cooling, we model the radiative cooling by molecules, which enables us to extend our analysis to lower densities.  We calculated the gas temperature using the code presented in \citet{Sipila18} and \citet{Sipila19}, which combines hydrodynamics with chemical and radiative transfer models in a self-consistent way.  For the present work, we assumed that the core has a static physical structure, so that the gas temperature is determined by balancing CR heating with cooling by the gas-dust collisional coupling and molecular line radiation.
 \par
\begin{figure}
\centering
\includegraphics[width=1.0\columnwidth]{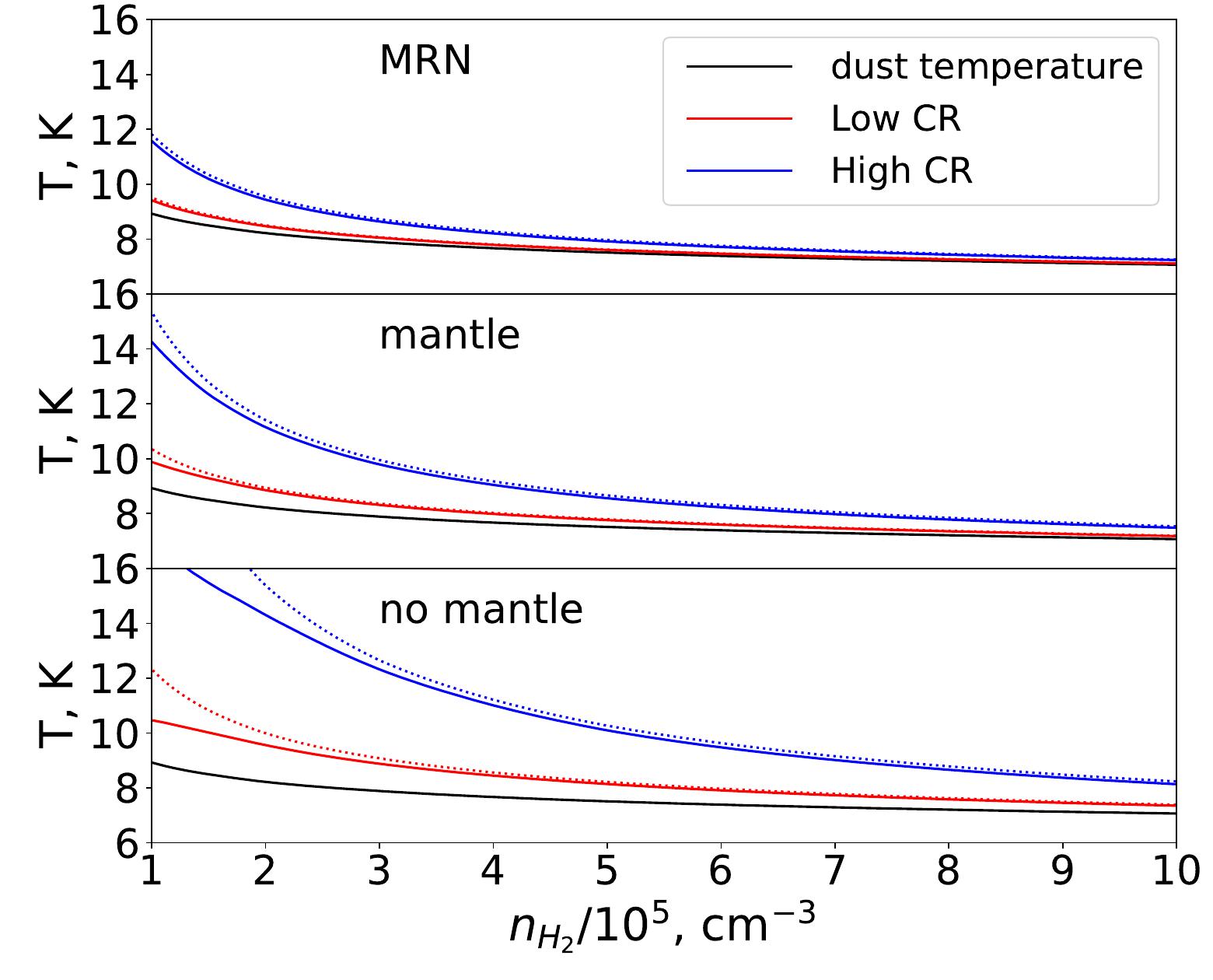}
\caption{Gas temperature as a function of position within the cloud.  Each panel corresponds to a different dust coagulation model, as labeled.  The black curve shows the dust temperature, as determined from the radiative transfer model.  The blue and red curves show the gas temperature assuming the high and low CR ionization rates, respectively, from \citet{Padovani18}.  The solid lines are the full cooling model, and the dashed lines correspond to the dust-only cooling model in \citet{Ivlev19}.  }
\label{fig:finalTemperature}
\end{figure}

 Figure \ref{fig:finalTemperature} shows the gas temperature as a function of local density in our model cloud.  This is made for three different dust coagulation models, as labeled on the panels.  The top panel shows results for an MRN size distribution with no mantles.  In the bottom two panels, the size distribution is determined by a coagulation simulation in which we assume compact spherical grains, no turbulence, and $c_d = 1$.  The middle panel shows a model in which grains have mantles as discussed in Section \ref{sect:mantles}.  The bottom panel is the same as the middle, but we have ignored the presence of icy mantles.  In each panel, the black line shows the dust temperature determined by the radiative transfer code CRT \citep{Juvela05}.  The curves labeled ``High CR'' and ``Low CR'' correspond to the high and low models in \citet{Padovani18}, which have respective  ionization rates of around $10^{-16}$ 
s$^{-1}$ and $2 \times 10^{-17}$ s$^{-1}$  at these densities.  
We did not run different dust coagulation simulations for the high- and low-$\zeta$ cases, but simply used different models for $\zeta$ to calculate the gas temperature.  The dotted lines correspond to temperatures from Equation 18 from \citet{Ivlev19}, and the solid lines are the results when line cooling is included as well. 
 Observations of gas temperatures in dense cores suggest that at densities
  of $10^5$ cm$^{-3}$ 
 the temperature is not higher than 12 Kelvin
  \citep{Crapsi07, Pagani07}.
 If true, this 
 constraint 
 would eliminate 
 both of our evolved dust
  models if the
   high
    CR rate 
 were accurate.  
 \par
 At higher densities the curves converge, and the differences between them become comparable to the uncertainty in the dust temperature.   At a density of $10^6$ cm$^{-3}$, \citet{Crapsi07} find a gas temperature of around 6\,K for the prestellar core L1544, and \citet{Pagani07} estimate 7\,K for the prestellar core L183.  Our model dust temperature at this density is 7.1K.  The warmest gas temperature of any of our models is only 8.1K.  It is worth noting that there is some degree of uncertainty in the dust temperature since it relies on a knowledge  of the optical properties of the absorbing dust and  on the strength of the IR field impinging on the cloud.  For these reasons, we believe the gas temperature at a density between 1 and 3 times $10^5$ cm$^{-3}$ has more ability to constrain dust growth and the CR rates than the temperatures at higher density.  

\subsection{Removal of small grains}
The question of the effect of the grain-size distribution on the ambipolar diffusion rate is important for more than just grain coagulation.  At higher densities, the ionization fraction drops furthexr, and small dust grains become very important for the coupling between the magnetic field and the neutral gas.  It was shown in \citet{Zhao16, Zhao18b} that removing the grains with sizes of less than a few tens of nanometers could increase both the ambipolar and Hall diffusivity by a factor of 10-100.  In simulations, this difference in the diffusivity results in a reduction in the magnetic flux dragged into the central disk-forming region, as well as the magnetic braking efficiency, which allows the formation of disks.  Our results show that precisely the grains that are relatively well coupled to the magnetic field are removed in collisions with larger grains, thus increasing the magnetic diffusivities.
\par
\citet{Ciolek96} noted that because of their coupling to the magnetic field, small grains will not drift to the center as fast as larger ones.  This provides another reason to think that small grains would be underabundant in the core center.  They predicted a reduction by a factor of $\sim 10$ in the number of 10 nm or smaller grains in the center.  In the absence of fragmentation, when collisions with larger grains are taken into consideration, we find a near complete elimination of such grains (see Figure \ref{fig:core}).  \citet{Ormel09} consider a model of fragmentation and find that fragmentation only becomes an issue after the particles are several microns in size.
\par
We do not explicitly take the polycyclic aromatic hydrocarbon population into account in this work, since their charging is uncertain.  Due to their small size, they will be in gross violation of our assumption that the charging timescales are shorter than the stopping time and inverse gyrofrequency.  Nevertheless, we believe the results shown here are sufficient to argue that they will be rapidly removed, similar to, or even faster than the smallest silicate grains.  If they are predominantly charged, then they will be coupled to the magnetic field in a way similar to the smallest silicate grains, so they will be removed in collisions with larger grains at about the same rate as the smallest silicate grains we consider.  If they were neutral, then their removal should be even faster, as they would be coupled to the neutral gas, and thus be moving near speed $v_{\rm in}$ with respect to the predominantly negatively charged grains with size $\leq a_c$, which make up most of the surface area of the initial distribution.  So, we predict that polycyclic aromatic hydrocarbons are efficiently adsorbed onto the icy mantles of larger dust grains, enriching their chemistry.
\section{Conclusions}
In this paper we discussed the role of different mechanisms that lead to grain-grain collisions.  In particular, we drew attention to a mechanism not considered in previous coagulation studies,  that of differential grain drift due to ambipolar diffusion.  We showed that this is more efficient than turbulence in removing grains smaller than about 50 nm from the size distribution.  This mechanism is not, however, a significant player in growing micron-sized grains.  Turbulence remains the only mechanism we are aware of that can greatly change the upper bound of the size distribution.  Turbulence is expected to be minimal near the core center, but may be present in the envelope.  Assuming perfect sticking in collisions, and relative velocities coming from ambipolar diffusion, turbulence, gravitational settling, and Brownian motion, we ran a series of coagulation simulations.  We reached the following general conclusions:
\begin{itemize}

\item{The effect of grain charge on the collision cross section due to Coulomb focussing-repulsion is not significant at densities of $10^6$ and below as there are mechanisms that lead to superthermal grain motion.}
\item{An increased CR ionization rate increases the core lifetime against ambipolar diffusion, thus resulting in larger grains if significant turbulence is present.  If the main growth mechanism is ambipolar diffusion, the differences are minimal as the increased lifetime is largely cancelled by the correspondingly slower differential drift of dust grains.}
\item{The inclusion of icy mantles, and the recognition that the non-spherical shape of grains significantly increases their collision cross section, has a significant effect on the degree of grain growth expected within the cloud lifetime.  With these effects considered, growth proceeds to 10 microns within the lifetime of the core if sonic turbulence is present at the core center.  Even without turbulence, essentially all grains under 100 nanometers are eliminated, but grains larger than a micron do not form. }  
\item{The resulting size distributions are nearly independent of the initial distribution of mantle thickness with grain size, but depend only on the total volume of mantle material relative to grain core, averaged over the size distribution.}
\item{ Particularly at densities above $10^6$ cm$^{-3}$, the removal of small grains can increase the ambipolar and Hall conductivities by over an order of magnitude.  This  greatly alleviates the magnetic braking problem which hinders disk formation in ideal magnetohydrodynamics.}
\item{
The gas temperature is a function both of the CR ionization rate and the dust grain-size distribution.  By carefully modeling the gas heating and cooling, we were able to show that the observed gas temperature near the center of dense cores such as L1544 is inconsistent with a CR ionization rate of $10^{-16}$ s$^{-1}$ if the grain growth expected from our ambipolar diffusion model has taken place, even without additional relative dust motion due to turbulence. }
\end{itemize}
The authors wish to thank the referee, Satoshi Okuzumi, for his careful reading of the manuscript and insightful comments that considerably improved the paper.  K.S. would like to thank Vincent Guillet for stimulating discussions.  A.V.I. acknowledges support by the Russian Science Foundation (project 18-12- 00351).
\bibliographystyle{apj}
\bibliography{arXivVersion.bib}

\begin{thebibliography}{}
\expandafter\ifx\csname natexlab\endcsname\relax\def\natexlab#1{#1}\fi

\bibitem[{{Asplund} {et~al.}(2009){Asplund}, {Grevesse}, {Sauval}, \&
  {Scott}}]{Asplund09}
{Asplund}, M., {Grevesse}, N., {Sauval}, A.~J., \& {Scott}, P. 2009, \araa, 47,
  481

\bibitem[{{Blum} \& {Wurm}(2008)}]{Blum08}
{Blum}, J., \& {Wurm}, G. 2008, \araa, 46, 21

\bibitem[{{Caselli} {et~al.}(2003){Caselli}, {van der Tak}, {Ceccarelli}, \&
  {Bacmann}}]{Caselli03}
{Caselli}, P., {van der Tak}, F.~F.~S., {Ceccarelli}, C., \& {Bacmann}, A.
  2003, \aap, 403, L37

\bibitem[{{Caselli} {et~al.}(1999){Caselli}, {Walmsley}, {Tafalla}, {Dore}, \&
  {Myers}}]{Caselli99}
{Caselli}, P., {Walmsley}, C.~M., {Tafalla}, M., {Dore}, L., \& {Myers}, P.~C.
  1999, \apjl, 523, L165

\bibitem[{{Caselli} {et~al.}(2002){Caselli}, {Walmsley}, {Zucconi}, {Tafalla},
  {Dore}, \& {Myers}}]{Caselli02}
{Caselli}, P., {Walmsley}, C.~M., {Zucconi}, A., {et~al.} 2002, \apj, 565, 344

\bibitem[{{Ciolek} \& {Basu}(2000)}]{Ciolek00}
{Ciolek}, G.~E., \& {Basu}, S. 2000, \apj, 529, 925

\bibitem[{{Ciolek} \& {Mouschovias}(1996)}]{Ciolek96}
{Ciolek}, G.~E., \& {Mouschovias}, T.~C. 1996, \apj, 468, 749

\bibitem[{{Crapsi} {et~al.}(2005){Crapsi}, {Caselli}, {Walmsley}, {Myers},
  {Tafalla}, {Lee}, \& {Bourke}}]{Crapsi05}
{Crapsi}, A., {Caselli}, P., {Walmsley}, C.~M., {et~al.} 2005, \apj, 619, 379

\bibitem[{{Crapsi} {et~al.}(2007){Crapsi}, {Caselli}, {Walmsley}, \&
  {Tafalla}}]{Crapsi07}
{Crapsi}, A., {Caselli}, P., {Walmsley}, M.~C., \& {Tafalla}, M. 2007, \aap,
  470, 221

\bibitem[{{Downes}(2012)}]{Downes12}
{Downes}, T.~P. 2012, \mnras, 425, 2277

\bibitem[{{Draine} \& {Sutin}(1987)}]{Draine87}
{Draine}, B.~T., \& {Sutin}, B. 1987, \apj, 320, 803

\bibitem[{{Fuller} \& {Myers}(1992)}]{Fuller92}
{Fuller}, G.~A., \& {Myers}, P.~C. 1992, \apj, 384, 523

\bibitem[{{Glassgold} {et~al.}(2012){Glassgold}, {Galli}, \&
  {Padovani}}]{Glassgold12}
{Glassgold}, A.~E., {Galli}, D., \& {Padovani}, M. 2012, \apj, 756, 157

\bibitem[{{Gredel} {et~al.}(1989){Gredel}, {Lepp}, {Dalgarno}, \&
  {Herbst}}]{Gredel89}
{Gredel}, R., {Lepp}, S., {Dalgarno}, A., \& {Herbst}, E. 1989, \apj, 347, 289

\bibitem[{{Gundlach} \& {Blum}(2015)}]{Gundlach15}
{Gundlach}, B., \& {Blum}, J. 2015, \apj, 798, 34

\bibitem[{{Ivlev} {et~al.}(2015){Ivlev}, {Padovani}, {Galli}, \&
  {Caselli}}]{Ivlev15}
{Ivlev}, A.~V., {Padovani}, M., {Galli}, D., \& {Caselli}, P. 2015, \apj, 812,
  135

\bibitem[{{Ivlev} {et~al.}(2019){Ivlev}, {Silsbee}, {Sipil{\"a}}, \&
  {Caselli}}]{Ivlev19}
{Ivlev}, A.~V., {Silsbee}, K., {Sipil{\"a}}, O., \& {Caselli}, P. 2019, \apj,
  884, 176

\bibitem[{{Juvela}(2005)}]{Juvela05}
{Juvela}, M. 2005, \aap, 440, 531

\bibitem[{{Keto} \& {Caselli}(2010)}]{Keto10}
{Keto}, E., \& {Caselli}, P. 2010, \mnras, 402, 1625

\bibitem[{{Kimura} {et~al.}(2015){Kimura}, {Wada}, {Senshu}, \&
  {Kobayashi}}]{Kimura15}
{Kimura}, H., {Wada}, K., {Senshu}, H., \& {Kobayashi}, H. 2015, \apj, 812, 67

\bibitem[{{K{\"o}nyves} {et~al.}(2015){K{\"o}nyves}, {Andr{\'e}},
  {Men'shchikov}, {Palmeirim}, {Arzoumanian}, {Schneider}, {Roy}, {Didelon},
  {Maury}, {Shimajiri}, {Di Francesco}, {Bontemps}, {Peretto}, {Benedettini},
  {Bernard}, {Elia}, {Griffin}, {Hill}, {Kirk}, {Ladjelate}, {Marsh}, {Martin},
  {Motte}, {Nguy{\^e}n Luong}, {Pezzuto}, {Roussel}, {Rygl}, {Sadavoy},
  {Schisano}, {Spinoglio}, {Ward-Thompson}, \& {White}}]{Konyves15}
{K{\"o}nyves}, V., {Andr{\'e}}, P., {Men'shchikov}, A., {et~al.} 2015, \aap,
  584, A91

\bibitem[{{Lada} {et~al.}(2008){Lada}, {Muench}, {Rathborne}, {Alves}, \&
  {Lombardi}}]{Lada08}
{Lada}, C.~J., {Muench}, A.~A., {Rathborne}, J., {Alves}, J.~F., \& {Lombardi},
  M. 2008, \apj, 672, 410

\bibitem[{{Lee} \& {Myers}(1999)}]{Lee99}
{Lee}, C.~W., \& {Myers}, P.~C. 1999, \apjs, 123, 233

\bibitem[{{Leger} {et~al.}(1985){Leger}, {Jura}, \& {Omont}}]{Leger85}
{Leger}, A., {Jura}, M., \& {Omont}, A. 1985, \aap, 144, 147

\bibitem[{{Mathis} {et~al.}(1977){Mathis}, {Rumpl}, \& {Nordsieck}}]{Mathis77}
{Mathis}, J.~S., {Rumpl}, W., \& {Nordsieck}, K.~H. 1977, The Astrophysical
  Journal, 217, 425

\bibitem[{{Musiolik} {et~al.}(2016){Musiolik}, {Teiser}, {Jankowski}, \&
  {Wurm}}]{Musiolik16}
{Musiolik}, G., {Teiser}, J., {Jankowski}, T., \& {Wurm}, G. 2016, \apj, 818,
  16

\bibitem[{{Nakamura} {et~al.}(2019){Nakamura}, {Kameno}, {Kusune}, {Mizuno},
  {Dobashi}, {Shimoikura}, \& {Taniguchi}}]{Nakamura19}
{Nakamura}, F., {Kameno}, S., {Kusune}, T., {et~al.} 2019, arXiv e-prints,
  arXiv:1908.07708

\bibitem[{{Neufeld} \& {Wolfire}(2017)}]{Neufeld17}
{Neufeld}, D.~A., \& {Wolfire}, M.~G. 2017, \apj, 845, 163

\bibitem[{{Okuzumi}(2009)}]{Okuzumi09}
{Okuzumi}, S. 2009, \apj, 698, 1122

\bibitem[{{Ormel} \& {Cuzzi}(2007)}]{Ormel07}
{Ormel}, C.~W., \& {Cuzzi}, J.~N. 2007, \aap, 466, 413

\bibitem[{{Ormel} {et~al.}(2009){Ormel}, {Paszun}, {Dominik}, \&
  {Tielens}}]{Ormel09}
{Ormel}, C.~W., {Paszun}, D., {Dominik}, C., \& {Tielens}, A.~G.~G.~M. 2009,
  \aap, 502, 845

\bibitem[{{Ossenkopf}(1993)}]{Ossenkopf93}
{Ossenkopf}, V. 1993, \aap, 280, 617

\bibitem[{{Padovani} {et~al.}(2018){Padovani}, {Ivlev}, {Galli}, \&
  {Caselli}}]{Padovani18}
{Padovani}, M., {Ivlev}, A.~V., {Galli}, D., \& {Caselli}, P. 2018, Astronomy
  and Astrophysics, 614, A111

\bibitem[{{Pagani} {et~al.}(2007){Pagani}, {Bacmann}, {Cabrit}, \&
  {Vastel}}]{Pagani07}
{Pagani}, L., {Bacmann}, A., {Cabrit}, S., \& {Vastel}, C. 2007, \aap, 467, 179

\bibitem[{{Pagani} {et~al.}(2010){Pagani}, {Steinacker}, {Bacmann}, {Stutz}, \&
  {Henning}}]{Pagani10}
{Pagani}, L., {Steinacker}, J., {Bacmann}, A., {Stutz}, A., \& {Henning}, T.
  2010, Science, 329, 1622

\bibitem[{{Pineda} {et~al.}(2010){Pineda}, {Goodman}, {Arce}, {Caselli},
  {Foster}, {Myers}, \& {Rosolowsky}}]{Pineda10}
{Pineda}, J.~E., {Goodman}, A.~A., {Arce}, H.~G., {et~al.} 2010, \apjl, 712,
  L116

\bibitem[{{Prasad} \& {Tarafdar}(1983)}]{Prasad83}
{Prasad}, S.~S., \& {Tarafdar}, S.~P. 1983, \apj, 267, 603

\bibitem[{Raizer {et~al.}(2011)Raizer, Kisin, \& Allen}]{Raizer11}
Raizer, Y., Kisin, V., \& Allen, J. 2011, Gas Discharge Physics (Springer
  Berlin Heidelberg)

\bibitem[{{Shen} {et~al.}(2008){Shen}, {Draine}, \& {Johnson}}]{Shen08}
{Shen}, Y., {Draine}, B.~T., \& {Johnson}, E.~T. 2008, \apj, 689, 260

\bibitem[{{Sipil{\"a}} \& {Caselli}(2018)}]{Sipila18}
{Sipil{\"a}}, O., \& {Caselli}, P. 2018, \aap, 615, A15

\bibitem[{{Sipil{\"a}} {et~al.}(2019){Sipil{\"a}}, {Caselli}, {Redaelli},
  {Juvela}, \& {Bizzocchi}}]{Sipila19}
{Sipil{\"a}}, O., {Caselli}, P., {Redaelli}, E., {Juvela}, M., \& {Bizzocchi},
  L. 2019, \mnras, 487, 1269

\bibitem[{{Tafalla} {et~al.}(2002){Tafalla}, {Myers}, {Caselli}, {Walmsley}, \&
  {Comito}}]{Tafalla02}
{Tafalla}, M., {Myers}, P.~C., {Caselli}, P., {Walmsley}, C.~M., \& {Comito},
  C. 2002, \apj, 569, 815

\bibitem[{{Tassis} \& {Mouschovias}(2004)}]{Tassis04}
{Tassis}, K., \& {Mouschovias}, T.~C. 2004, \apj, 616, 283

\bibitem[{{Wada} {et~al.}(2013){Wada}, {Tanaka}, {Okuzumi}, {Kobayashi},
  {Suyama}, {Kimura}, \& {Yamamoto}}]{Wada13}
{Wada}, K., {Tanaka}, H., {Okuzumi}, S., {et~al.} 2013, \aap, 559, A62

\bibitem[{{Weingartner} \& {Draine}(2001)}]{Weingartner01}
{Weingartner}, J.~C., \& {Draine}, B.~T. 2001, \apj, 548, 296

\bibitem[{{Xu} {et~al.}(2016){Xu}, {Yan}, \& {Lazarian}}]{Xu16}
{Xu}, S., {Yan}, H., \& {Lazarian}, A. 2016, \apj, 826, 166

\bibitem[{{Zhao} {et~al.}(2018{\natexlab{a}}){Zhao}, {Caselli}, \&
  {Li}}]{Zhao18b}
{Zhao}, B., {Caselli}, P., \& {Li}, Z.-Y. 2018{\natexlab{a}}, \mnras, 478, 2723

\bibitem[{{Zhao} {et~al.}(2018{\natexlab{b}}){Zhao}, {Caselli}, {Li}, \&
  {Krasnopolsky}}]{Zhao18a}
{Zhao}, B., {Caselli}, P., {Li}, Z.-Y., \& {Krasnopolsky}, R.
  2018{\natexlab{b}}, \mnras, 473, 4868

\bibitem[{{Zhao} {et~al.}(2016){Zhao}, {Caselli}, {Li}, {Krasnopolsky},
  {Shang}, \& {Nakamura}}]{Zhao16}
{Zhao}, B., {Caselli}, P., {Li}, Z.-Y., {et~al.} 2016, \mnras, 460, 2050

\end{thebibliography}
\end{document}